\documentclass[sigconf,table]{acmart}

\setcopyright{none} 
\acmConference[Anonymous Submission to ACM CCS 2019]{ACM Conference on Computer and Communications Security}{Due 15 May 2019}{London, TBD}
\acmYear{2019}

\usepackage{soul}
\usepackage{marginnote}
\usepackage{listings}
\usepackage{microtype}
\usepackage[most]{tcolorbox}
\usepackage{lipsum}
\usepackage{xfrac}
\usepackage{pifont}
\usepackage{array}
\hypersetup{
    colorlinks=true,
    linkcolor=blue!60!black,
    citecolor=red!60!black,
    urlcolor=blue!80!black,
    filecolor=magenta!60!black,
    menucolor=red!60!black,
    anchorcolor=black,
    pdfstartview=FitH,
    breaklinks=true
}

\usepackage{graphicx}
\usepackage{colortbl}
\usepackage{multirow}
\usepackage{hyperref}
\usepackage{amsmath}
\usepackage{pifont}
\usepackage{array}

\newcolumntype{C}[1]{>{\centering\arraybackslash}p{#1}}

\definecolor{mediumgreen}{rgb}{0.0, 0.75, 0.0}

\lstdefinestyle{mypython}{
  language=Python,
  basicstyle=\ttfamily\small,
  keywordstyle=\color{blue},
  commentstyle=\color{gray},
  stringstyle=\color{orange},
  showstringspaces=false,
  breaklines=true,
}

\definecolor{SummaryBg}{HTML}{E3F2FD}
\definecolor{SummaryFrame}{HTML}{1976D2}
\newtcolorbox{summarybox}{
  colback=SummaryBg,
  colframe=SummaryFrame,
  fonttitle=\bfseries,
  title=Summary,
  boxrule=1pt,
  arc=2mm,
  left=3pt, right=3pt, top=3pt, bottom=3pt,
  enlarge left by=0mm,
  enlarge right by=0mm
}

\begin{document}
\title{EinHops: Einsum Notation for Expressive \\Homomorphic Operations on RNS-CKKS Tensors} 

\author{Karthik Garimella}
\affiliation{
  \institution{New York University}
  \city{}
  \country{}
}
\email{kg2383@nyu.edu}

\author{Austin Ebel}
\affiliation{
  \institution{New York University}
  \city{}
  \country{}
}
\email{abe5240@nyu.edu}

\author{Brandon Reagen}
\affiliation{
  \institution{New York University}
  \city{}
  \country{}
}
\email{bjr5@nyu.edu}

\begin{abstract}
Fully Homomorphic Encryption (FHE) is an encryption scheme that allows for computation to be performed directly on encrypted data. FHE effectively closes the loop on secure and outsourced computing; data is encrypted not only during rest and transit, but also during processing. Moreover, modern FHE schemes such as RNS-CKKS (with the canonical slot encoding) encrypt one-dimensional floating-point vectors, which makes such a scheme an ideal candidate for building private machine learning systems. However, RNS-CKKS provides a limited instruction set: SIMD addition, SIMD multiplication, and cyclic rotation of these encrypted vectors. This restriction makes performing multi-dimensional tensor operations (such as those used in machine learning) challenging. Practitioners must pack multi-dimensional tensors into 1-D vectors and map tensor operations onto this one-dimensional layout rather than their traditional nested structure. And while prior systems have made significant strides in automating this process, they often hide critical packing decisions behind layers of abstraction, making debugging, optimizing, and building on top of these systems difficult. 

In this work we ask: can we build an FHE tensor system with a straightforward and transparent packing strategy regardless of the tensor operation? We answer affirmatively and develop a packing strategy based on Einstein summation (einsum) notation. We find einsum notation to be ideal for our approach since the notation itself explicitly encodes the dimensional structure and operation directly into its syntax, naturally exposing how tensors should be packed and manipulated in FHE. We make use of einsum's explicit language to decompose einsum expressions into a fixed set of FHE-friendly operations: dimension expanding and broadcasting, element-wise multiplication, and a reduction along the contraction dimensions. 

We implement our design and present EinHops, which stands for \underline{Ein}sum Notation for \underline{H}omomorphic Tensor \underline{Op}erations. EinHops is a minimalist system that factors einsum expressions into a fixed sequence of FHE operations, enabling developers to perform complex tensor operations using RNS-CKKS while maintaining full visibility into the underlying packing strategy. We evaluate EinHops on a range of tensor operations from a simple transpose to complex multi-dimensional contractions. We show that the explicit nature of einsum notation allows us to build an FHE tensor system that is simple, general, and interpretable. We open-source EinHops at the following repository: \href{https://github.com/baahl-nyu/einhops}{https://github.com/baahl-nyu/einhops}.  

\end{abstract}




\maketitle

\section{Introduction}
\label{sec:intro}

Fully Homomorphic Encryption (FHE) is a powerful encryption scheme that enables computation to be performed directly on encrypted data without the need for decryption~\cite{fhe}. This property makes FHE a promising fit for outsourced computation, especially for applications that have strict data privacy rules (e.g., health care). Within FHE, there are several schemes that operate over different data types: TFHE/CGGI for encrypted boolean circuits~\cite{tfhe}, BFV/BGV for arithmetic circuits over encrypted integer vectors~\cite{b,fv,bgv}, and CKKS for arithmetic circuits over encrypted complex-valued (and therefore real-valued) vectors~\cite{ckks}. For this reason, the CKKS scheme (in particular, RNS-CKKS~\cite{rns-ckks}) has been used as the backbone for developing outsourced privacy-preserving machine learning services ~\cite{orion, fhelipe, chet, hecate, helayers}. These systems execute encrypted deep learning applications by performing tensor operations within the constraints of RNS-CKKS.

\begin{figure}[t]
\vspace{0.3cm}
\begin{lstlisting}[
    style=mypython,
    backgroundcolor=\color{gray!10},
    frame=single,
    frameround=tttt,
    rulecolor=\color{blue!30},
    numbers=left,
    numberstyle=\tiny\color{gray},
    commentstyle=\color{green!60!black}\itshape,
    keywordstyle=\color{blue!80!black}\bfseries,
    stringstyle=\color{red!70!black},
    captionpos=b,
    caption={EinHops introduces einsum notation~\cite{pytorch-einsum} to the RNS-CKKS FHE scheme~\cite{rns-ckks}. This example performs batched matrix multiplication where both operands are encrypted.},
    label={lst:overview}
]
import torch
import einhops

# batched matrix multiplication (pytorch)
a = torch.randn(2, 3, 4)
b = torch.randn(2, 4, 5)
c = torch.einsum("bij,bjk->bik", a, b)

# encrypted batched matrix multiplication (ckks)
a_ctxt = einhops.encrypt(a)
b_ctxt = einhops.encrypt(b)
c_ctxt = einhops.einsum("bij,bjk->bik", 
                        a_ctxt, 
                        b_ctxt)

# verify correctness
assert c.shape == c_ctxt.shape == (2, 3, 5)
assert torch.allclose(c, einhops.decrypt(c_ctxt))
\end{lstlisting}
\end{figure}

When implementing encrypted multi-dimensional tensor operations in RNS-CKKS, we must first map our multi-dimensional data onto 1-D vectors. This issue is also present in canonical tensor libraries such as PyTorch or Jax in which multi-dimensional tensors are mapped to blocks of memory (i.e., the logical representation versus the physical representation)~\cite{pytorch,jax,yang2019pytorchinternals}. The key difference when applying this mapping problem to RNS-CKKS is the limited instruction set provided by FHE: we only have access to SIMD (Single Instruction, Multiple Data) addition, SIMD multiplication, and cyclic rotations~\cite{rns-ckks}. And unlike PyTorch, which can freely index into memory, FHE cannot access individual elements without rotating the entire vector. This makes indexing or accessing encrypted sub-tensors fundamentally more challenging.

Prior systems like Fhelipe and CHET~\cite{fhelipe, chet} have made significant strides in bridging this gap through sophisticated compiler infrastructures that automatically handle packing, bootstrapping, and optimization. However, these approaches often obscure the crucial packing decisions behind layers of abstraction. For instance, when developers use a high-level function like matvec or matmul, they lose visibility into whether their vectors are packed row-wise or column-wise, whether padding is used, or how reductions are implemented. This abstraction is appropriate for modern deep learning libraries that implement such operations using highly optimized kernels (e.g., cuDNN) \cite{cudnn, cuBLAS, BLAS}. On the other hand, linear algebra subroutines in FHE are still being developed without a consensus on the optimal packing strategy, given that the underlying cost model is inherently different. In FHE, we must ask: How many multiplicative levels does your method consume? What encoding procedure is being used? How many unique rotation keys are required? Can you use hoisted rotations? How much memory is required?

In this work, rather than building a large-scale system, we choose to build a simple FHE tensor system guided by the principle that the packing strategy and FHE costs should be explicit to the end-user. We find einsum notation to be an ideal candidate for our approach since it naturally exposes the dimensional structure of tensor operations through its notation. We implement our design in EinHops, which brings the simple but powerful einsum notation to the RNS-CKKS FHE scheme~. Listing \ref{lst:overview} shows how einsum notation can be used to perform multi-dimensional tensor operations by explicitly labeling dimensions over which to perform contractions. EinHops runs the equivalent operations on encrypted tensors.

EinHops implements this design philosophy by recognizing a key structural equivalence between einsum notation and a sequence of FHE operations. For example, the reduction in \texttt{"ij,jk->ik"} maps naturally to a sequence of operations in the RNS-CKKS slot space: first, a linear transformation to align the data, a sequence of rotations and summations to replicate data, an element-wise multiplication, and finally another sequence of rotations and additions to perform the reduction over the contracted dimension.

Concretely, we make the following contributions:
\begin{itemize}
    \item We develop a transparent packing strategy based on einsum notation that exposes data layout decisions and remains consistent across arbitrary tensor operations.
    \item We implement and open-source our system, EinHops, with support for both CPU and GPU through an existing FHE backend, plus a cleartext backend for slot-level debugging.
    \item We evaluate EinHops on more than 15 tensor operations, showing that explicit packing enables operations such as 5-D tensor contractions while remaining interpretable.
\end{itemize}

The rest of the paper is organized as follows. Section \ref{sec:background} provides the relevant background information on the RNS-CKKS FHE scheme and einsum notation. In Section \ref{sec:decomp_einsum}, we decompose einsum expressions into a series of explicit FHE-friendly operations, and in Section \ref{sec:einhops} we implement each of these steps using FHE primitives. We discuss the system design and limitations of EinHops in Section \ref{sec:system}, and we report our results in Section \ref{sec:eval}.

\section{Background}
\label{sec:background}

\subsection{RNS-CKKS}
\begin{figure}
\centering
\includegraphics[width=\columnwidth]{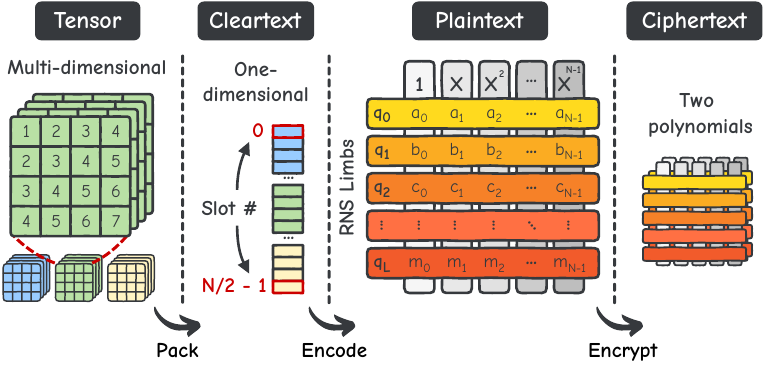}
\caption{RNS-CKKS encrypts one-dimensional vectors of a fixed length (e.g., $2^{15}$ slots) and enables SIMD Addition, SIMD multiplication, and cyclic rotation upon these encrypted vectors. Multi-dimensional tensors must therefore be \textit{packed} into one-dimensional cleartext vectors. These cleartext vectors are then \textit{encoded} into plaintext polynomials that are decomposed using RNS~\cite{rns}. Finally, plaintexts can be \textit{encrypted} into ciphertexts, which consist of two RNS-polynomials.}
\label{fig:data_structures}
\end{figure}

Here, we describe the RNS-CKKS homomorphic encryption scheme \cite{rns-ckks}, its relevant datatypes, and the pipeline that consists of data packing, encoding, and encryption. A high level overview of this process is shown in Figure \ref{fig:data_structures}.

\subsubsection{Datatypes} We interact with four primary datatypes in CKKS: tensors, cleartexts, plaintexts, and ciphertexts. Tensors are the familiar multi-dimensional arrays that represent our input data in its logical, high-level form and exist outside of the scope of FHE. A cleartext is a one-dimensional vector of a fixed power-of-two length (e.g., $2^{15}$ slots). We pack tensors into one or more cleartexts. A plaintext is then generated by encoding a cleartext vector into a single polynomial that resides in the ring $\mathcal{R}_Q = \mathbb{Z}_Q[X]/(X^N + 1)$, where $N$ is a power-of-two degree (e.g., $2^{16}$)  and $Q$ is a large integer modulus. Encryption then converts a plaintext into a ciphertext, which consists of a pair of  polynomials in the product ring $\mathcal{R}_Q \times \mathcal{R}_Q$.

\subsubsection{Packing} The process of flattening multi-dimensional tensors into one-dimensional cleartext vectors is called packing. Techniques for packing have been extensively explored in prior work \cite{orion, fhelipe, chet, lee2022} and the best solutions are often problem-dependent. For example, packing strategies in convolutional neural networks \cite{orion} generally differ from those in language models \cite{thor} due to their distinct network architectures. 

\subsubsection{Encoding} The goal of encoding in CKKS is to find a polynomial in the ring $\mathcal{R}_Q$ whose evaluations at $\sfrac{N}{2}$ complex-valued roots of unity interpolate our cleartext vector. This process involves applying a variant of the inverse discrete Fourier transform. The conversion from point-value representation to coefficient representation guarantees that additions and multiplications between polynomials perform element-wise, SIMD operations on their underlying cleartexts. In this way, we can \text{pack} $\sfrac{N}{2}$ values into the \textit{slots} of our cleartext vector.
To satisfy RLWE constraints, a polynomial’s coefficients must also be integers. Therefore before interpolating, we scale the evaluations by a large factor $\Delta$ (typically a power of two) to embed the desired precision into the integer parts. Then, rounding to the nearest integer incurs less error. Decoding performs the inverse transformation to recover scaled evaluations of any resultant cleartext. We note that other encoding schemes such as coefficient encoding exist~\cite{coeff-encoding}, but do not preserve the SIMD properties we would like to utilize in this work.

\subsubsection{Encryption} We can encrypt the scaled integer plaintext polynomial $m(X)$ into a ciphertext $ct = (b(X), a(X))$. The component $a(X)$ is typically sampled uniformly at random from the ring $\mathcal{R}_{Q}$. The second component, $b(X)$, is then constructed based on $a(X)$, the secret key $s(X)$ (a small polynomial in $\mathcal{R}_Q$), the plaintext $m(X)$, and a freshly sampled small error polynomial $e(X)$. Specifically, $b(X)$ is computed as $[-a(X) \cdot s(X) + m(X) + e(X)]_{Q}$, where the operations are performed with integer coefficients before the final reduction modulo $Q$. This structure securely masks $m(X)$.
Decryption then uses the secret key $s(X)$ to reverse this masking. The decryption process computes $m'(X) = [b(X) + a(X) \cdot s(X)]_{Q}$. Substituting the definition of $b(X)$, we see the simplification:
\begin{align*}
m'(X) &\equiv b(X) + a(X) \cdot s(X) \pmod{Q} \\
      &\equiv (-a(X) \cdot s(X) + m(X) + e(X)) + a(X) \cdot s(X) \pmod{Q} \\
      &\equiv m(X) + e(X) \pmod{Q}
\end{align*}
As long as the coefficients of the combined $m(X)+e(X)$ are small enough, this final $m'(X)$ correctly recovers the original scaled plaintext plus the initial encryption error.

\subsubsection{Homomorphic Operations} CKKS allows for computations directly on ciphertexts, corresponding to operations on the underlying cleartext vectors. The scheme supports element-wise addition and multiplication of these vectors. If $ct_1$ encrypts a vector corresponding to $m_1(X)$ and $ct_2$ encrypts a vector for $m_2(X)$, their homomorphic sum, $ct_{\text{add}} = ct_1 \oplus ct_2$, decrypts to approximately $m_1(X)+m_2(X)$, effectively adding the slot values. The same property holds for homomorphic multiplication. Additionally, CKKS supports homomorphic rotations (cyclic shifts) of the encrypted vector of slot values. This is achieved by applying a specific Galois automorphism to the ciphertext, which permutes the underlying data within the slots without decryption. All homomorphic operations are performed in the ring $R_{Q}$, where the ciphertext modulus $Q$ must be very large to accommodate for noise growth.

\subsubsection{Residual Number System (RNS)} Working directly with the large integer modulus $Q$ is compute intensive. To address this, CKKS implementations adopt the Residue Number System (RNS)~\cite{rns}. The RNS approach decomposes the single large modulus $Q$ into a product of several smaller, typically machine-word-sized, pairwise coprime moduli $q_0, q_1, \ldots, q_L$, so that $Q = Q_L = \prod_{i=0}^{L} q_i$. The integer $L$ is known as the maximum multiplicative $\textit{level}$ of the ciphertext. By the Chinese Remainder Theorem, any integer coefficient $c$ from a CKKS polynomial (which is an element of $\mathbb{Z}_{Q_L}$) can then be uniquely represented by a vector of its residues $(c \bmod{q_0}, \ldots, c \bmod{q_L})$. Thus each polynomial in $\mathcal{R}_{Q_L}$ is transformed into a tuple of "limb" polynomials, where the $i$-th limb has coefficients modulo the small prime $q_i$. Operations originally performed on the large-coefficient polynomials are now efficiently computed through independent, parallel operations on their corresponding small-coefficient polynomial limbs using standard machine-word arithmetic.

\subsubsection{Level Consumption} Homomorphic multiplication in CKKS consumes levels of the RNS modulus. This occurs because multiplication significantly increases both the magnitude of the scaled plaintext (from a scaling factor of $\Delta$ in the inputs to $\Delta^2$ in the product before rescaling) and the accumulated noise. To manage these increases, \textit{rescaling} is performed after multiplication. Rescaling effectively divides the ciphertext by the original scaling factor $\Delta$ (or an RNS prime $q_j$ that approximates it) and, crucially, reduces the ciphertext modulus from the current $Q_l = \prod_{i=0}^{l} q_i$ to a smaller $Q_{l-1} = \prod_{i=0}^{l-1} q_i$ by dropping one of the RNS primes (e.g., $q_l$). Each rescaling operation therefore \textit{consumes} one RNS modulus or one level. Since fewer limbs exist at lower levels, the latency of homomorphic operations increases with increasing level. When no further levels can be consumed, an expensive bootstrapping operation must be performed to refresh this multiplicative budget. As a result, the latency of bootstrapping often dominates the runtime of practical workloads such as private deep neural network inference.

\subsection{Einsum Notation}
\label{sec:einsum}
\noindent
\subsubsection{Overview}
Einsum notation is a language for performing tensor operations by explicitly stating over which dimensions to perform contractions~\cite{einstein}. This notation is a simple, but powerful abstraction that allows one to succinctly express a variety of tensor operations such as transposes, matrix-vector products, matrix-matrix multiplications, and higher-dimension tensor contractions.

While originally used in physics, einsum notation has since been adopted in modern tensor libraries such as PyTorch and Jax~\cite{pytorch, jax}, and each of these libraries is equipped with an \texttt{einsum} function. This function takes as input 1) an \texttt{equation} that specifies the tensor operation and 2) a series of operands. As an example, the expression \texttt{torch.einsum("ij->ji", x)} from Line 3 of Listing \ref{listing:einsum_examples} tells us that we have a 2-D input matrix \texttt{x} with shape \texttt{"ij"}. Furthermore, the equation specifies that the resulting output tensor should have shape \texttt{"ji"}, informing us that we are transposing \texttt{x}.

By explicitly labeling both the input dimensions and the desired output dimensions, einsum notation facilitates tensor arithmetic while being \textit{interpretable}. We make use of einsum's explicitness to develop EinHops, a system that performs encrypted tensor operations without obscuring the packing and implementation details.

\subsubsection{Einsum Examples} Einsum notation is best understood by reading and running examples~\cite{rocktaschel2018einsum}. We begin with Listing \ref{listing:einsum_examples} which instantiates a random tensor \texttt{x} of size $(3, 5)$. Each of the \texttt{einsum} calls specifies the input tensor as having shape \texttt{"ij"}, effectively labeling the axis \texttt{"i"}$= 3$ and \texttt{"j"}$= 5$. We could not have used \texttt{einsum} with the input dimensions being listed as\texttt{"ijk"} given that the input tensor \texttt{x} is two-dimensional.

The first \texttt{einsum} call (Line 3) performs the transpose operation by switching the order of the dimensions in the output (labeled \texttt{"ji"}). The subsequent calls to \texttt{einsum} perform a reduction over one or more dimensions by omitting these dimensions from the output equation. For example, Line 4 reduces over both dimensions by having no output dimensions specified. On the other hand, Line 5 reduces over the \texttt{"i"} dimension by stating that the output dimension should only be \texttt{"j"}$= 5$.

\begin{lstlisting}[
    float=t,
    style=mypython,
    backgroundcolor=\color{gray!10},
    frame=single,
    frameround=tttt,
    rulecolor=\color{blue!30},
    numbers=left,
    numberstyle=\tiny\color{gray},
    commentstyle=\color{green!60!black}\itshape,
    keywordstyle=\color{blue!80!black}\bfseries,
    stringstyle=\color{red!70!black},
    captionpos=b,
    caption={Einsum notation to perform a transpose and various summations for a 2-D tensor illustrating that einsum explicitly labels each dimension from the inputs.},
    label={listing:einsum_examples}
]
import torch
x = torch.randn(3,5)
torch.einsum("ij->ji", x) # transpose       (5,3)
torch.einsum("ij->", x)   # sum all elements 
torch.einsum("ij->j", x)  # column-wise sum  (5,)
torch.einsum("ij->i", x)  # row-wise sum     (3,)
\end{lstlisting}

\begin{lstlisting}[
    style=mypython,
    backgroundcolor=\color{gray!10},
    frame=single,
    frameround=tttt,
    rulecolor=\color{blue!30},
    numbers=left,
    numberstyle=\tiny\color{gray},
    commentstyle=\color{green!60!black}\itshape,
    keywordstyle=\color{blue!80!black}\bfseries,
    stringstyle=\color{red!70!black},
    captionpos=b,
    caption={A matrix-vector product using einsum notation. In this case, the dimension \texttt{"j"} is multiplied and then reduced.},
    label={listing:einsum_matvec}
]
import torch
x = torch.randn(3,5)
y = torch.randn(5)

# matrix-vector product: (3,5) x (5,) -> (3,)
torch.einsum("ij,j->i", x, y)
\end{lstlisting}

\begin{figure*}[!t]
\centering
\includegraphics[width=\textwidth]{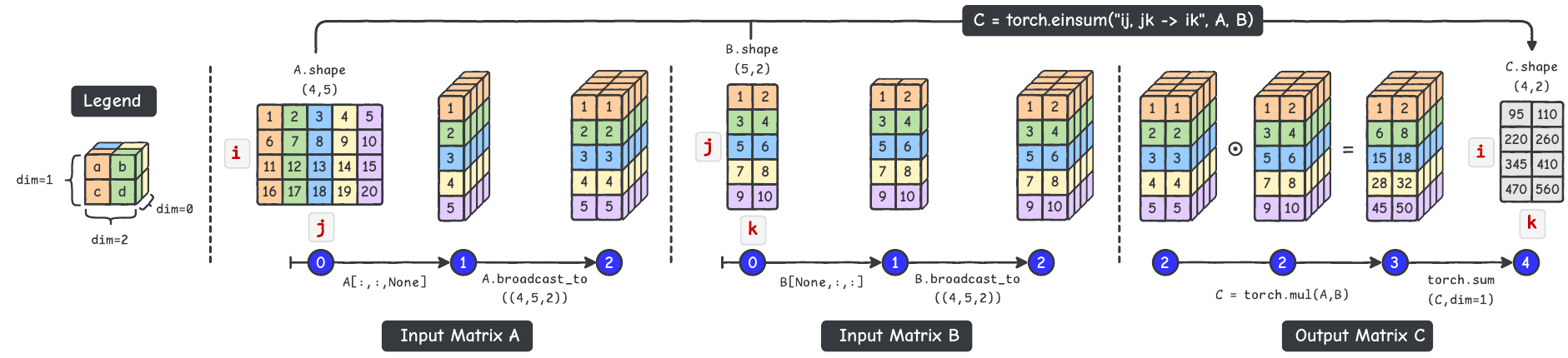}
\caption{Einsum notation for a matrix-matrix multiplication in \texttt{PyTorch} of size $(4,5) \times (5,2) \rightarrow (4,2)$. The einsum equation labels these input dimensions as $(i,j)\times(j,k)$ and explicitly states the output shape as $(i,k)$. The dataflow decomposes this einsum equation in an FHE-friendly manner. First, all operands are expanded and broadcasted to match tensor shapes. Then, operands are element-wise multiplied. Finally, the output is reduced by summing across the contraction dimensions.}
\label{fig:einsum}
\end{figure*}

Listing \ref{listing:einsum_matvec} has an example of an einsum expression with two inputs, \texttt{x} and \texttt{y}. The \texttt{equation} argument illustrates this by labeling each dimension of the inputs, separated by a comma. In this instance, 
we are performing a matrix-vector product. The \texttt{"j"} dimension is shared between both inputs and is omitted from the output dimensions. This tells us that we are performing a contraction over dimension \texttt{"j"}:  the inputs are multiplied and then reduced along the \texttt{"j"} dimension, effectively computing dot products over \texttt{"j"}.

\subsubsection{Einsum in Practice}
We now examine a real-world application of the einsum notation from a recent ICLR 2024 paper~\cite{deletang2024language}. In Listing \ref{listing:mha}, einsum notation appears in  key steps when performing the multi-headed attention mechanism. By explicitly labeling each dimension, the einsum expressions are self-documenting, making them translatable to natural language. For example, Line 5 of Listing \ref{listing:mha} can be read as: "for every sample in the batch (\texttt{b}) and for every head (\texttt{h}), construct the \texttt{t} by \texttt{T} attention matrix by contracting the inputs over their shared hidden dimension (\texttt{d})".

\begin{lstlisting}[
    style=mypython,
    backgroundcolor=\color{gray!10},
    frame=single,
    frameround=tttt,
    rulecolor=\color{blue!30},
    numbers=left,
    numberstyle=\tiny\color{gray},
    commentstyle=\color{green!60!black}\itshape,
    keywordstyle=\color{blue!80!black}\bfseries,
    stringstyle=\color{red!70!black},
    captionpos=b,
    float, % <--- ADD THIS OPTION
    caption={Since input and output dimensions are explicitly labeled, einsum notation is self-documenting as shown in this \href{https://github.com/google-deepmind/language_modeling_is_compression/blob/main/transformer.py\#L88}{implementation} of multi-headed attention.},
    label={listing:mha}
]
import jax.nn as jnn
import jax.numpy as jnp

# batch_size, seq_len, n_heads, h_dim_per_head
attn = jnp.einsum("bthd,bThd->bhtT", q, k)
n_attn = jnn.softmax(attn)
output = jnp.einsum("bhtT,bThd->bthd", n_attn, v)
\end{lstlisting}

\section{Decomposing Einsum for FHE (\textit{\small\MakeLowercase{concept}})}

\label{sec:decomp_einsum}

In this section, we \textit{examine} einsum notation from the perspective of FHE. In particular, we develop a mapping from einsum expressions to three FHE-friendly steps: 1) expanding and broadcasting inputs to match dimensions, 2) performing an element-wise multiplication between the broadcasted inputs, and 3) reducing the resulting product across the contraction dimensions.

To clarify, standard einsum expressions are \textbf{\textit{not} }implemented using these three steps; in reality, an \texttt{einsum} call will dispatch to a highly optimized backend implementation that depends upon the size, sparsity, and device associated with each input~\cite{opt_einsum, einsum_bench}. Rather, our conceptual view of einsum lets us carry over the explicit dimensionality analysis into the realm of FHE by building a system that explicitly realizes this decomposition. We use Listing \ref{listing:decomp} and Figure \ref{fig:einsum} as our reference for this section, which performs a matrix-matrix multiplication between two 2-D tensors.

\subsection{Matching Dimensions (\texttt{torch})}
First, we analyze the shapes of each input as well as the total number of input dimensions. In our running example, the einsum expression tells us that the input $A$ has shape \texttt{"i"}$=4$ and \texttt{"j"}$=5$, whereas the input $B$ has shape \texttt{"j"}$=5$ and \texttt{"k"}$=2$. The total number of input dimensions is three; we have \texttt{\{ "i", "j", "k"\}}.

Our first FHE-friendly step is to expand and match the shapes of the inputs. This means we must inflate $A$ to include the missing input dimension \texttt{"k"} and inflate $B$ to include the missing input dimension \texttt{"i"}. We label this Step \ding{192} and Step \ding{193} in Figure \ref{fig:einsum}.

\subsubsection{Expanding}
We first match the \textit{number} of input dimensions by simply adding singleton dimensions in an ordering that satisfies \href{https://github.com/srush/Tensor-Puzzles?tab=readme-ov-file#rules}{broadcasting rules}~\cite{pytorch_broadcasting}. In Figure \ref{fig:einsum}, we insert these new axes using \texttt{A[:, :, None]} and \texttt{B[None, :, :]} although we could have also used \texttt{view}, \texttt{reshape}, or \texttt{unsqueeze}.

The shapes of $A$ and $B$ are now $(4,5,1)$ and (1,5,2) as shown in Step \ding{192} in Figure \ref{fig:einsum}. Note that we comply with broadcasting rules and that the contraction dimension (\texttt{"j"}$=5$) is aligned at  the \texttt{dim=1} position. This step performs no data duplication. Rather, it simply \textit{views} the tensor within a higher dimension. As we will see in the following section, we can perform this operation in FHE either as a linear transformation or a \texttt{nop}, depending upon where the singleton dimensions must be added.

\subsubsection{Broadcasting}
Now that their shapes are aligned, we can broadcast both inputs $A$ and $B$ to match their sizes along each dimension. In this step, we duplicate the data across the singleton dimensions; we visualize this duplication in Step \ding{193} of Figure \ref{fig:einsum} where the tensors $A$ and $B$ now have shapes $(4, 5, 2)$. In the following section, we will show how to perform this broadcasting across any dimension in FHE using a logarithmic number of operations.

\subsection{Multiplication (\texttt{torch})}
Now that the input tensor shapes precisely match up, we may perform an element-wise Hadamard product between the inputs $A$ and $B$. Here, we are effectively performing all partial products in the matrix-matrix multiplication in parallel, and this results in an output tensor $C$ of shape $(4, 5, 2)$. This resulting tensor $C$ is shown in Step \ding{194} of Figure \ref{fig:einsum}. We note that this multiplication naturally extends to the scenario when we have more than two inputs since all input tensors will have the same shape. 
\begin{lstlisting}[
    style=mypython,
    backgroundcolor=\color{gray!10},
    frame=single,
    frameround=tttt,
    rulecolor=\color{blue!30},
    numbers=left,
    numberstyle=\tiny\color{gray},
    commentstyle=\color{green!60!black}\itshape,
    keywordstyle=\color{blue!80!black}\bfseries,
    stringstyle=\color{red!70!black},
    captionpos=b,
    float, 
    caption={Matrix-matrix multiplication using \texttt{torch.einsum}.},
    label={listing:decomp}
]
import torch
A = torch.arange(start=1,end=21).reshape(4,5)
B = torch.arange(start=1,end=11).reshape(5,2)
C = torch.einsum("ij,jk->ik", A, B)
\end{lstlisting}
\subsection{Reduction (\texttt{torch})}
Finally, we must reduce all partial products by summing over the contraction dimension, which in this case is \texttt{"j"}$=5$. This step can be seen in Step \ding{195} of Figure \ref{fig:einsum} where we sum out \texttt{dim=1} which corresponds to the labeled dimension \texttt{"j"}, resulting in a tensor of shape $(4, 2)$. This resulting tensor precisely corresponds to the desired matrix-matrix multiplication performed in Listing \ref{listing:decomp}. Similar to broadcasting, we will show how to perform this reduction over the contraction dimensions in a logarithmic number of FHE operations in the following section. 

\begin{figure*}[!t]
\centering
\includegraphics[width=\textwidth]{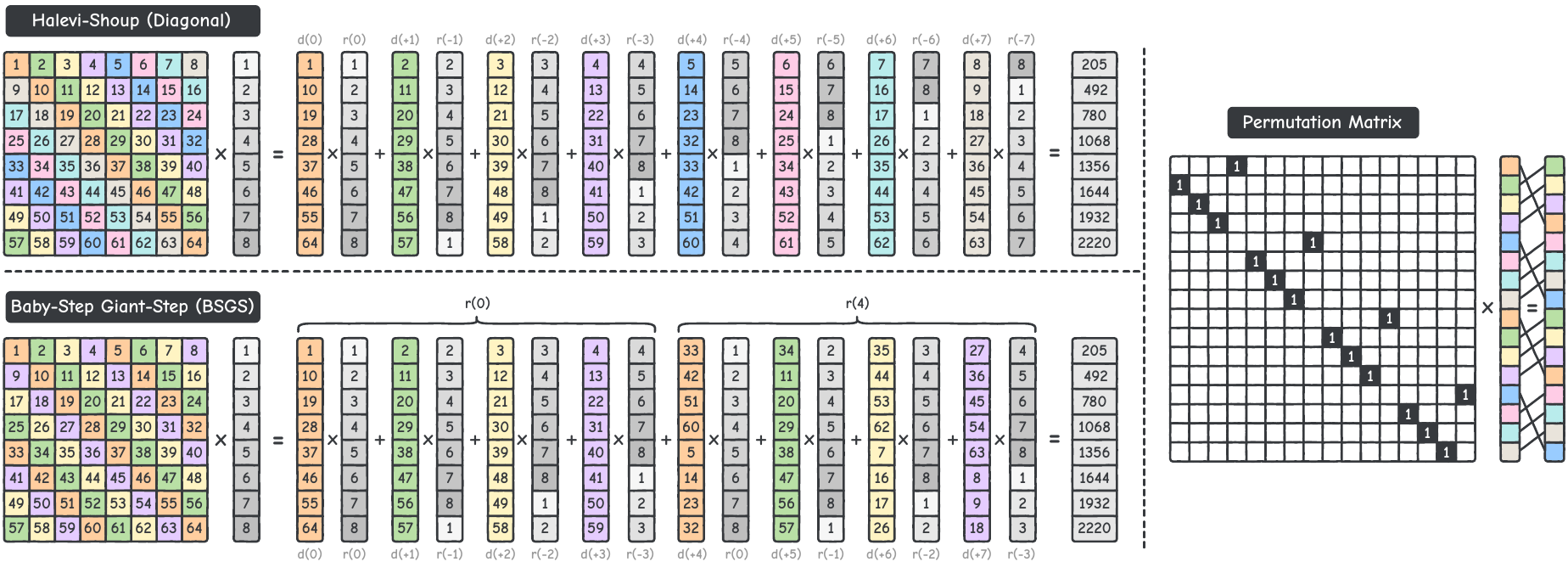}
\caption{(\textit{Left}) Matrix-vector products between an unencrypted matrix and an encrypted vector with $n$ slots. The canonical Halevi-Shoup method multiplies the diagonals of the matrix with the (homomorphically) aligned ciphertext~\cite{halevishoup}. For a full $n\times n$ matrix, Halevi-Shoup requires $n$ homomorphic rotations. The Baby-Step Giant-Step (BSGS) algorithm pre-rotates the cleartext diagonals, therefore reducing the homomorphic rotation count to  $O(\sqrt{n})$~\cite{bootstrapping}. (\textit{Right}) EinHops utilizes BSGS to re-arrange the slots of an encrypted vector by performing a matrix-vector product between a permutation matrix and a ciphertext.}
\label{fig:bsgs_rotsum}
\end{figure*}

\section{Decomposing Einsum for FHE (\textit{\small\MakeLowercase{impl.}})}
\label{sec:einhops}

We now \textit{implement} our decomposed view of einsum from the previous section using RNS-CKKS primitives. As a consequence, we follow the same outline from the previous section and show the explicit set of FHE instructions that carry out 1) expanding and broadcasting, 2) element-wise multiplication, and 3) summation along the contraction dimensions. We assume that the product of all input dimensions fit within a single ciphertext, and that all dimensions are padded up to powers of two. 

For this section, we demonstrate our low-level implementation details in Figures \ref{fig:bsgs_rotsum}, \ref{fig:broadcast}, and \ref{fig:reduction}. At the end of this section, we use Figure \ref{fig:einhops} to illustrate the end-to-end EinHops implementation of the same matrix-matrix multiplication from Listing \ref{listing:decomp} and Figure \ref{fig:einsum}.

\subsection{Matching Dimensions (FHE)}
As we saw in the previous section, we must first expand our tensors by inserting singleton dimensions in order to match the \textit{number} of inputs dimensions for each operand. In  libraries such as PyTorch or Jax, such an expansion is trivial and for tensors allocated in contiguous memory blocks, these expansions only change tensor metadata without modifying the actual data itself~\cite{yang2019pytorchinternals}. 

However for encrypted tensors, there is a possibility that we must actually manipulate the ciphertext slot ordering given that indexing-based tensor operations are invalid (i.e. we only have SIMD Add, SIMD Mult, and cyclic rotation). There are two possible scenarios for dimension expansion which we detail below. To re-iterate, we are not duplicating data during expansion; we are merely re-ordering data within the slots of a CKKS ciphertext.

\subsubsection{Expanding inner dimensions}
When inserting an inner dimension into an encrypted tensor, we must permute the underlying elements within the slots of the ciphertext. This is because inserting an inner dimension will change the \textit{stride} of existing dimensions when the tensor is flattened~\cite{pytorch_stride}. To see why, let us examine the input tensor in the bottom of Figure \ref{fig:broadcast} (\texttt{input.shape == (4,1)}). In this case, we must insert an inner dimension: $(4, 1) \rightarrow (4,2)$. Before any expansion, the data within the input ciphertext is contiguously located in the top four slots of the underlying vector. However, adding an inner dimension will naturally stride the original elements, in this case by a stride of $2$. This means we must insert placeholder slots for this new dimension between the original elements, and we do this by permuting the slots of the CKKS ciphertext using a linear transformation.

In more detail, we choose to perform these permutations as a linear transformation via a permutation matrix. Figure \ref{fig:bsgs_rotsum} (left) shows the Baby-Step Giant-Step (BSGS) algorithm that we use to perform this linear transformation. The BSGS algorithm is a derivative of the more straightforward Halevi-Shoup matrix-vector product algorithm. Halevi-Shoup multiplies the plaintext diagonals of a matrix with a homomorphically rotated ciphertext, effectively sharding each dot product across the aligned slots of many ciphertexts as shown. Concretely, the first row of the matrix performs a dot product with the ciphertext elements across slot index 0, the second row with slot index 1, and so forth. BSGS provides precisely the same interface as Halevi-Shoup but reduces the homomorphic rotation count from $O(n)$ to $O \sqrt{n}$ by instead rotating (i.e. \texttt{torch.roll}) the cleartext diagonals before encoding them into plaintexts. Figure \ref{fig:bsgs_rotsum} (right) shows that the diagonal structure of a permutation matrix fits naturally with BSGS.

\subsubsection{Expanding outer dimensions}
Expanding outer dimensions is a much simpler scenario since the original input data remains in contiguous slots. An example of this setup is shown in the top of Figure \ref{fig:broadcast} (\texttt{input.shape == (1,4)}) where we must add an outer dimension: $(1, 4) \rightarrow (2, 4)$. Here, the original input data retains its stride of 1, so no permutation matrix is required to prepare this tensor for broadcasting. This effectively becomes a \texttt{nop} in FHE since the data is already in the correct order.

\subsubsection{Broadcasting}
\begin{figure}
\centering
\includegraphics[width=0.9\columnwidth]{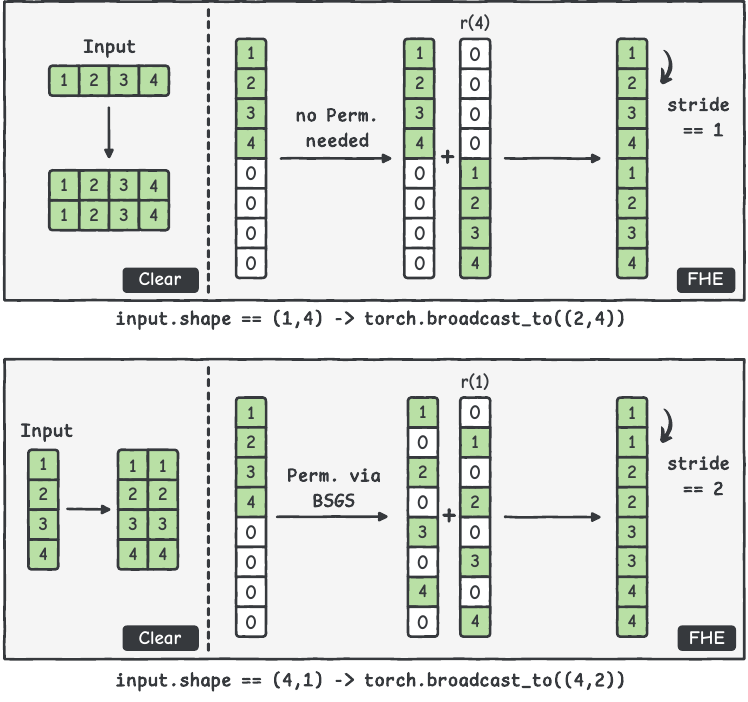}
\caption{Broadcasting in FHE using $O(\log n)$ operations. (\textit{Top}) Expanding  outer dimensions ($(1,4) \rightarrow (2,4)$) does not permute the ordering of existing elements. (\textit{Bottom}) Expanding inner dimensions ($(4,1) \rightarrow (4,2)$) strides existing elements and requires a permutation before broadcasting.}
\label{fig:broadcast}
\end{figure}

At this stage, the input tensors have been expanded by inserting placeholder slots via BSGS in the case of inner dimension expansion or performing a \texttt{nop} in the case of outer dimension expansion. We are now ready to broadcast our original data across the new dimensions as shown in Figure \ref{fig:broadcast}. 

Because we have chosen to pad all dimensions to a power of 2, we can perform broadcasting across any set of dimensions using the rotation-and-summation algorithm down the slots of the ciphertext while only requiring a logarithmic number of homomorphic operations with respect to the dimension size~\cite{heco}. The \textit{number} of homomorphic rotations is logarithmic in the \textit{size} of the new dimension. The number of \textit{places} by which we rotate the vector depends upon the \textit{stride} of the new dimension. In both the top and bottom of Figure \ref{fig:broadcast}, we only perform a single homomorphic rotation because the size of the new dimension is 2 and so $\log (2) = 1$. In the top of Figure \ref{fig:broadcast}, we rotate down by four since the stride of the new dimension is $4$. In the bottom, we rotate down by one since the stride of the new dimension is $1$.

\subsection{Multiplication (FHE)}
Now that all ciphertext tensors have been broadcasted using FHE primitives and have matching dimensions, it is straightforward to perform the multiplication using homomorphic SIMD Multiplication. For $k$ operands, we can perform this step in $O(\log k)$ multiplicative levels by performing a tree-based multiplication.

\subsection{Reduction (FHE)}
\begin{figure}
\centering
\includegraphics[width=0.9\columnwidth]{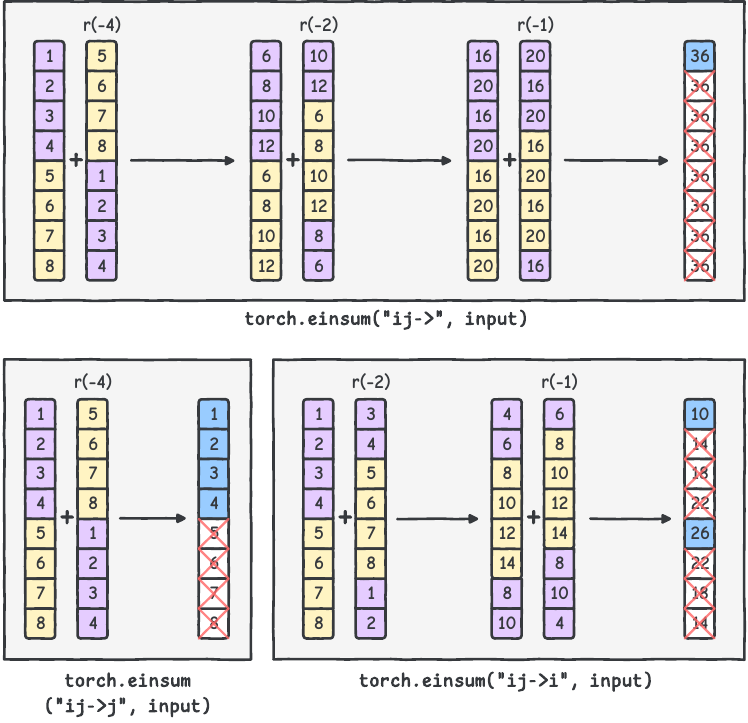}
\caption{Reductions in FHE using $O(\log n)$ operations. Summing an inner dimension \texttt{("ij->i")} introduces gaps between the desired output, whereas summing the outer dimension \texttt{("ij->j")} produces the desired sums in contiguous slots.}
\label{fig:reduction}
\end{figure}

Now, we must reduce the output ciphertext across the contraction dimensions. Similar to broadcasting, we can perform reductions across any arbitrary set of dimensions by using rotation-and-summation up the slots of the ciphertext. We illustrate how to perform this reduction in Figure \ref{fig:reduction} for a 2-D tensor. Again, we need a logarithmic number of rotations with respect to the sizes of the contraction dimensions, and the number of places by which we rotate is dictated by the strides of the contraction dimensions.

Here, we make a critical observation that facilitates a simpler FHE implementation. When performing a reduction over an inner dimension (e.g., \texttt{torch.einsum("ij->i", input)}) as shown in the bottom right of Figure \ref{fig:reduction}, the desired sums are strided by the size of this inner dimension. This introduces gaps in the output ciphertext. On the other hand, reducing over an outer dimension (e.g., \texttt{torch.einsum("ij->j", input)}) as shown in the bottom left of Figure \ref{fig:reduction} naturally places the output sums in contiguous slots at the beginning of the ciphertext.

We use this observation to make sure that the resulting reduction places the computation in contiguous slots at the top of the ciphertext. In more detail, when we parse our einsum equation in EinHops to build our expanded and broadasted dimensions, we always place the contraction dimensions as the outer dimensions and the desired output dimensions as the inner dimensions in the correct order. This design choice enables us to forgo any permutation to re-align the output into contiguous slots in the output ciphertext.

\begin{figure*}[!t]
\centering
\includegraphics[width=\textwidth]{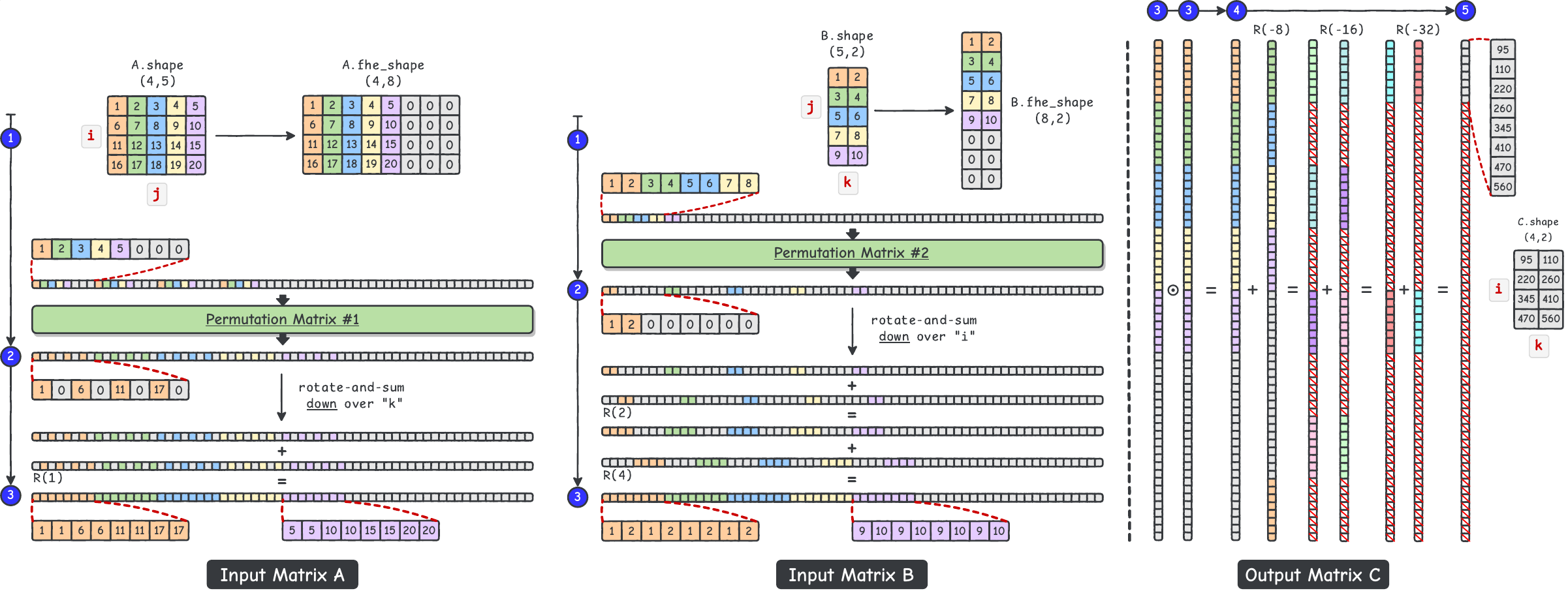} 
\caption{An overview of EinHops using the same example from Listing \ref{listing:decomp} and Figure \ref{fig:einsum}. Step \ding{192} the einsum equation is parsed in order to determine the broadcast shape. Step \ding{193} if required, we permute the encrypted inputs using a BSGS permutation. Step \ding{194} each input is broadcasted to match tensor shapes. Step \ding{195} all inputs are multiplied using SIMD multiplication. Step \ding{196} we perform a reduction to collapse the contraction dimensions. Step \ding{197} a final mask ensures that all unused slots are zeroed out.}
\label{fig:einhops}
\end{figure*}

\subsection{Putting it all Together}
We now use Figure \ref{fig:einhops} to illustrate this entire process of 1) expanding and broadcasting input dimensions, 2) performing the element-wise multiplication, and 3) reducing along the contraction dimensions for the same matrix-matrix multiplication example from Listing \ref{listing:decomp}.
\subsubsection{Inputs}
The inputs $A$ and $B$ have shapes $(4,5)$ and $(5,2)$, respectively. These are padded up to the nearest power of two to $(4,8)$ and $(8,2)$, but we keep the original dimensions as an attribute for both tensors. We can now encrypt these two padded tensors. The equation \texttt{"ij,jk->ik"} in Listing \ref{listing:decomp} informs us that we are reducing over dimension \texttt{"j"} and desire an output dimension of \texttt{"ik"}. We choose our broadcasted dimension to be \texttt{"jik"} in order to place the contraction dimension as the outer dimension and the resulting dimensions as inner dimensions in the correct order.

\subsubsection{Matching}
The inputs $A$ and $B$ have dimensions \texttt{"ij"} and \texttt{"jk"}, respectively. Expanding both to have a dimension of \texttt{"jik"} requires inserting an inner dimension and so we need permutations for both inputs. Furthermore, the permutation for $A$ includes the transpose from \texttt{"ij"} to \texttt{"ji"}. The permutations (Step \ding{193} in Figure \ref{fig:einhops}) result in ciphertexts with placeholder slots for broadcasting. We now perform broadcasting using the rotation-and-summation algorithm to ensure that we fill up the first \texttt{"jik"} slots in each ciphertext with the correct and aligned data. This process is shown as Step \ding{194} of Figure \ref{fig:einhops} and results in tightly packed ciphertexts.

\subsubsection{Multiplying}
We can now perform a simple SIMD multiplication using our backend FHE library. This will perform all the partial products required by the matrix-multiplication and result in a ciphertext tensor with the same dimension (\texttt{"jik"}) as shown in Step \ding{195} of Figure \ref{fig:einhops}.

\subsubsection{Reducing}
We now reduce over the contraction dimension \texttt{"j"}. We know that the (FHE) size of \texttt{"j"} is $8$ and the stride of \texttt{"j"} is also $8$. So, we perform the logarithmic rotation-and-summation algorithm up the ciphertext. This requires $\log(8) = 3$ steps and the places by which we rotate starts at $8$ (the stride) and doubles each time. Since we placed the contraction dimension as the outer dimension, we've calculated our desired matrix-matrix multiplication in the top contiguous slots of the ciphertext. The output after this reduction is shown in Step \ding{196} of Figure \ref{fig:einhops}. We will perform a final masking of all other slots to maintain correctness for any further computations (i.e. multiplication by a binary vector).

\section{EinHops System}
\label{sec:system}

In this section, we discuss the overall design philosophy, implementation details, and limitations of EinHops. Our code is open-sourced at: \href{https://github.com/baahl-nyu/einhops}{https://github.com/baahl-nyu/einhops}.

\subsection{Design Philosophy} 
At a high level, the theme of EinHops is \textit{minimalism} and \textit{simplicity}. This means that we would like to be explicit about our packing strategy and we prefer eager execution versus graph execution which is the de-facto standard in FHE tensor systems~\cite{orion, fhelipe, chet}. This design choice has ramifications in terms of raw performance but greatly simplifies the EinHops architecture. 

\subsubsection{Goals} Similar to PyTorch or Jax, we would like to provide an \texttt{einsum} interface for expressing tensor reductions and contractions in FHE. Inputs can either be tensors or encrypted tensors. We target small FHE programs such as tensor contractions, matrix-vector products, and matrix-matrix multiplication. Finally, we would like to provide a simple and readable code-base to provide a scaffolding for future research and engineering. 

\subsubsection{Non-goals} We do not support running large FHE programs (e.g., encrypted ResNet-20 or LLM inference). Running such programs requires a robust system that handles the sharper bits of FHE (e.g., memory management, bootstrap placement, scale management, data normalization, and non-linear approximations)~\cite{orion}. 

\subsection{Implementation}
\begin{figure}
\centering
\includegraphics[width=\columnwidth]{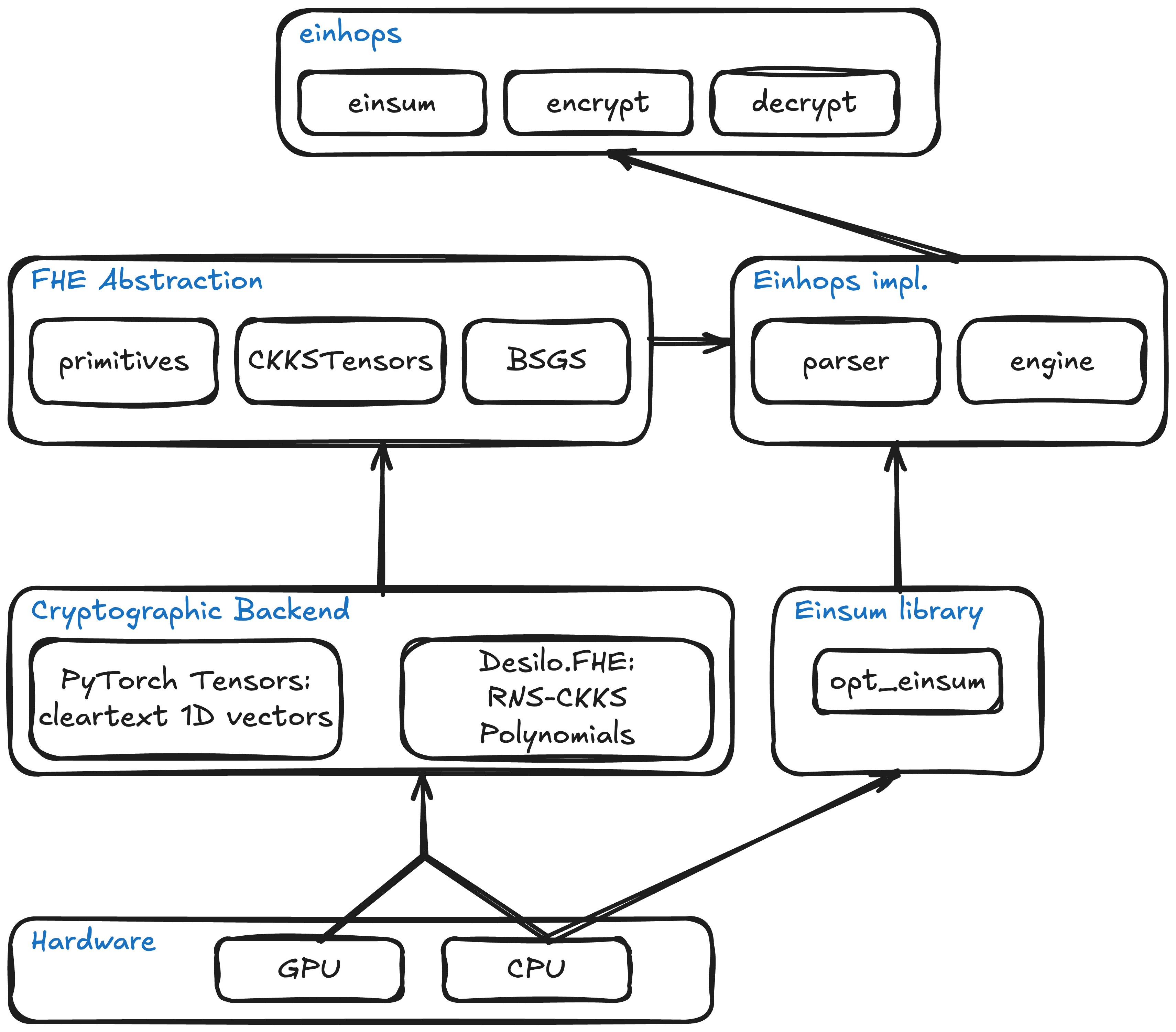}
\caption{EinHops systems overview from the hardware to user interface. We implement this system in pure \texttt{Python}.}
\label{fig:system}
\end{figure}

Figure \ref{fig:system} shows the overall system architecture for EinHops, which we implement in roughly 1,000 lines of pure \texttt{Python} code. 
\subsubsection{Backend}
We currently support the \texttt{Liberate.FHE} library by using their \texttt{desilo-fhe} \texttt{Python} frontend~\cite{Liberate_FHE}. This library lowers FHE operations to both CPUs and GPUs. Additionally, we support an artificial backend of PyTorch 1-D cleartext tensors. This means that we can simulate homomorphic operations using \texttt{torch.add}, \texttt{torch.mul}, and \texttt{torch.roll}, which brings debugging FHE programs to the slot level with minimal overhead. We use the \texttt{opt\_einsum} package to parse and validate einsum expressions.

\subsubsection{Hoisted Baby-Step Giant-Step}
Since the \texttt{desilo-fhe} backend only provides an interface for homomorphic SIMD addition, SIMD multiplication, and cyclic rotation on \texttt{desilo} objects, we implement the baby-step giant-step algorithm directly in \texttt{Python} with the single hoisting optimization that amortizes the cost of key-switching within each baby step~\cite{faster_helib}.

\subsection{Limitations}
Here, we briefly describe the current limitations of EinHops.

\setlength{\tabcolsep}{2.70pt}
\setlength{\arrayrulewidth}{0.25mm}
\renewcommand{\arraystretch}{1.3}
\begin{table*}[t]
    \small
   \centering
   \begin{tabular}{|C{3.65cm}|C{3.65cm}|C{4.0cm}|C{1.25cm}|C{1.25cm}|C{1.25cm}|C{1.25cm}|}
    \hline
    \cellcolor{gray!20}\textbf{Operation} & 
    \cellcolor{gray!20}\textbf{Syntax} & 
    \cellcolor{gray!20}\textbf{Input Dimensions} & 
    \cellcolor{gray!20}\textbf{FHE CPU  (s)} & 
    \cellcolor{gray!20}\textbf{FHE CPU + Keys (s)} & 
    \cellcolor{gray!20}\textbf{FHE GPU (s)} & 
    \cellcolor{gray!20}\textbf{FHE GPU + Keys (s)} \\

     \hline \noalign{\vskip 0.1cm}  \hline

     Matrix transpose       & $\mathtt{ij \rightarrow ji}$                      & $(128, 128)$                        & $60.55$  & $19.58 $ & $11.52$                      & $5.91$  \\ \cline{1-7}
     Matrix sum             & $\mathtt{ij}\rightarrow$                          & $(128, 128)$                        & $2.80$ & $2.09$ & $0.44$                        & $0.40$  \\ \cline{1-7}
     Column sum             & $\mathtt{ij \rightarrow j}$                       & $(128, 128)$                        & $0.77$ & $0.49$ & $0.12$                        & $0.11$  \\ \cline{1-7}
     Row sum                & $\mathtt{ij \rightarrow i}$                       & $(128, 128)$                        & $62.59$ & $19.83$ & $10.16$                       & $6.26$  \\ \cline{1-7}
     Matrix $\times$ vector & $\mathtt{ik,k \rightarrow i}$                     & $(128, 128), (128)$                 & $124.41$ & $34.84$ & $17.69$                & $11.58$ \\ \cline{1-7}
     Matrix $\times$ matrix & $\mathtt{ik,kj \rightarrow ij}$                   & $(16, 32), (32, 32)$                & $27.16$ & $10.17$ & $5.91$                & $4.97$  \\ \cline{1-7}
     Dot product            & $\mathtt{i,i \rightarrow}$                        & $(16384,)$                          & $2.50$ & $1.63$ & $0.27$                            & $0.33$  \\ \cline{1-7}
     Inner product          & $\mathtt{ij, ij\rightarrow}$                      & $(128, 128), (128,128)$             & $2.26$ & $1.84$  & $0.27$            & $0.31$  \\ \cline{1-7}
     Hadamard product       & $\mathtt{ij, ij\rightarrow ij}$                   & $(128, 128), (128,128)$             & $0.05$ & $0.03$  & $0.01$            & $0.01$ \\ \cline{1-7}
     Outer product          & $\mathtt{i, j\rightarrow ij}$                     & $(128,), (128,)$                    & $61.16$ & $16.21$ & $8.55$                    & $5.00$  \\ \cline{1-7}
     Batched matrix $\times$ matrix        & $\mathtt{ijk, ikl\rightarrow ijl}$ & $(16,8,8),(16,8,8)$                 & $11.54$ & $3.70$  & $4.12$               & $4.01$  \\ \cline{1-7}
     $3$-way Hadamard       & $\mathtt{ij,ij,ij \rightarrow ij}$                & $(128, 128), (128,128), (128, 128)$ & $0.08$ & $0.05$ & $0.01$ & $0.01$  \\ \cline{1-7}
     Chained matrix $\times$ matrix        & $\mathtt{ij,jk,kl \rightarrow il}$ & $(16,8),(8,8),(8,16)$               & $85.34$ & $21.31$  & $14.22$             & $9.30$  \\ \cline{1-7}
     Bilinear transform     & $\mathtt{ik, jkl, il\rightarrow ij }$             & $(8, 16), (8, 16, 16), (8, 16)$     & $167.89$ & $69.16$ & $29.2$   & $27.07$ \\ \cline{1-7}
     Tensor contraction     & $\mathtt{pqrs, tuqvr\rightarrow pstuv}$           & $(2,4,8,8), (1,4,4,2,8)$            & $104.28$ & $28.13$  & $16.12$           & $12.04$ \\ 

    \hline
   \end{tabular}
   \vspace{4px}
   \caption{EinHops on a CPU and GPU FHE backend. The \textbf{"+ Keys"} columns indicate the we generate the specific rotations keys needed for the BSGS matrix-vector product apriori. BSGS rotations keys increase working memory from $\approx 3$ GB to $\approx 32$GB, but reduce computation by enabling a single-step rotation rather than a series of power-of-two-step rotations for the BSGS stage.}
   \label{tab:your_new_label}
\end{table*}

\subsubsection{Single Ciphertext} Currently, EinHops only supports operations between single ciphertexts and as a consequence, the product of all input dimensions to an einsum expression must fit within a single ciphertext, effectively limiting our total dimension size to $N/2$. Adding support for multi-ciphertext tensors unlocks a new scaling dimension for EinHops. Given that EinHops builds wide computation graphs, task-based optimization can be used to increase throughput~\cite{cudastf}. At the same time, the multi-ciphertext case raises several crucial questions over the meaning of applying homomorphic operations across ciphertexts. For example: what does it mean to rotate a multi-ciphertext by some value $k$? Is this a global rotation by $k$ or a per-ciphertext rotation by $k$?

\subsubsection{Bootstrap Placement} We do not currently implement any bootstrapping placement policy. While it is straightforward to implement a lazy bootstrapping strategy, this method has shown to scale poorly, especially with arithmetic circuits that contain residual connections~\cite{orion}.

\subsubsection{Permutations}
We implement permutations right now as BSGS linear transformations. However, it would beneficial to employ a different permutation strategy such as the Vos-Vos-Erkin method~\cite{kun2024shiftnetworks, vve} and analyze the change in terms of required memory, latency, and level consumption.

\subsubsection{jit support}
Compiling a series of einsum calls in EinHops may reveal better packing and permutations strategies rather than employing eager execution mode.

\section{Evaluation}
\label{sec:eval}
In this section, we report our results for evaluating EinHops on a series of 15 einsum expressions~\cite{rocktaschel2018einsum} as well as calculating the multi-headed self-attention score (Line 5 of Listing \ref{listing:mha}). All experimental results are averaged over 10 runs and we observe a small standard deviation in each case. We also calculate the $\ell_2$ norm of the difference between the PyTorch and EinHops \texttt{einsum} expressions, and observe a discrepancy on the order of roughly $10^{-5}$ in each case. For every experiment, we choose to encrypt all inputs to the einsum expression, although this does not have to be the case. Indeed, the EinHops implementation of \texttt{einsum} accepts both torch tensors and ciphertexts.

\subsection{Experimental Setup} All experiments are performed on an Intel(R) Xeon(R) Platinum 8480+ processor (26 threads, 13 cores, 2 threads per core) that is equipped with a single NVIDIA H100 PCIe GPU (80GB memory). We run the FHE backend on both the CPU (using all 26 threads) and the GPU. We use the \href{https://fhe.desilo.dev/latest/create_engine/}{medium engine} provided by the \texttt{desilo-fhe} backend which has a polynomial degree of $2^{15}$ and a corresponding slot count of $16384$. We run each experiment at the minimum starting level: this is level $4$ for \texttt{Chained matrix $\times$ matrix} and \texttt{Bilinear transform} and level $3$ for all other einsum expressions.

For both the CPU and GPU backend, we also run two different configurations of EinHops. The first configuration is memory-efficient but requires more compute by generating only the power-of-two rotation keys (14 keys in our setup). This means, for example that rotating a ciphertext by 3 slots requires two distinct rotations: a rotation by 1 slot, followed by a rotation by 2 slots. The second variant (labeled \texttt{+ Keys} in Table \ref{tab:your_new_label}) is compute-efficient but requires more working memory by generating all rotation keys needed for a general BSGS matrix transformation (14 + 256 keys in our setup).

\subsection{Results} 
Table \ref{tab:your_new_label} lists our results for running 15 different einsum expressions using EinHops ~\cite{rocktaschel2018einsum} for both the CPU and GPU backend as well as our two key-generation configurations. We list the tensor operation along with the corresponding einsum syntax on the left side of the Table. For example, the \texttt{Chained matrix $\times$ matrix} operation performs two back-to-back matrix-matrix multiplications where all matrices are ciphertexts. For each experiment, we specify the input dimension so as to use a majority of the available slots in the CKKS ciphertext (16384 slots in our setup). 

\subsubsection{Power-of-Two Keys Only}
For the memory-efficient setup, which generates only the power-of-two rotation keys, we find that in most cases the GPU backend achieves roughly a $6 \times$ speedup compared to the CPU backend. For example, transposing a $128 \times 128$ matrix takes 60.55 seconds on the CPU backend while taking only 11.52 seconds on the GPU. In this particular instance, no reduction is occurring and all time is spent permuting the ciphertext slots using BSGS. Still, this memory-efficient configuration requires only 3 GB of working memory, which lowers the barrier to entry for running and understanding FHE programs. While we choose to run our experiments on server-grade hardware, this modest memory requirement means EinHops can run on consumer-grade equipment such as a laptop or an RTX 3090 GPU.

\subsubsection{Power-of-Two + BSGS Keys}
The compute-efficient setup, which generates both the power-of-two rotation keys and the BSGS rotation keys immediately lowers runtime in both the CPU and GPU setting. This is because we are able to perform the rotations with the pre-computed BSGS rotation keys rather than using a series of power-of-two steps~\cite{keyswitch}. Furthermore in this setup, we are able to leverage single-hoisting to amortize the Decompose component of rotations for the baby-step rotations~\cite{faster_helib}. As an example, the compute-efficient setup on the GPU results in a $10.7\times$ speedup to compute a \textit{ciphertext-ciphertext} matrix-vector product on inputs of size $(128 \times 128)$ and $(128,1)$ when compared to the CPU backend which generates only the power-of-two rotation keys. Still, this compute-efficient configuration requires roughly 32 GB of working memory to hold all power-of-two and BSGS rotation keys.

\subsubsection{Case Study: Multi-Headed Self Attention Scores} Besides the tensor operations from Table \ref{tab:your_new_label}, we also use EinHops to compute the multi-head attention score from Listing \ref{listing:mha} (Line 5). We set the batch size, sequence length, number of heads, and hidden dimension to be \texttt{b=2}, \texttt{t=T=5}, \texttt{h=8}, \texttt{d=16}, respectively. Again, both inputs to the attention mechanism are encrypted. We are able to compute the attention matrix for each sample and each head in 28.57 seconds with roughly 23 seconds being spent aligning data using the BSGS transformation. This experiment illustrates the importance of researching strategies for re-aligning data within CKKS slots. Overall, our results demonstrate that einsum notation provides a clean abstraction for FHE tensor operations without sacrificing the transparency of underlying slot manipulations.

\section{Related Works}
\label{sec:related}

The development of EinHops is informed by two primary areas of research: the evolution of compilers for FHE and the use of einsum notation in tensor libraries.

\vspace{2px}

\noindent \textbf{FHE Compilers:} The field of FHE compilation has rapidly matured to address the significant performance and programmability challenges of computing on encrypted data. Early works focused on circuit-level optimizations, and more recent domain-specific compilers such as CHET \cite{chet}, EVA \cite{eva}, Porcupine \cite{porcupine}, and Hecate \cite{hecate} targeted shallow machine learning workloads. These systems introduce abstractions to automate complex tasks such as encryption parameter selection, noise management through sophisticated rescaling strategies (e.g., EVA's "waterline"), and optimizing data layouts for SIMD execution. More advanced compilers such as Orion \cite{orion}, Dacapo \cite{dacapo}, and Fhelipe \cite{fhelipe} tackle the challenge of \textit{deep} learning by automating the placement of bootstrapping operations.

A common thread among prior FHE compilers is their reliance on generating static computational graphs. These systems typically parse a program into an intermediate representation and apply a series of optimization passes to manage noise, latency, and scheduling. We instead adopt an eager execution model akin to PyTorch \cite{pytorch}. Instead of building and optimizing a graph, the einsum expression itself serves as an explicit plan. This design prioritizes simplicity and transparency, making the packing strategy explicit rather than abstracting it away behind an optimization framework.

\vspace{2px}

\noindent \textbf{Einsum Notation:} Einsum notation has become a staple in modern tensor libraries such as PyTorch \cite{pytorch}, JAX \cite{jax}, and NumPy \cite{numpy} for its ability to express complex tensor operations in a single, concise line of code. The notation's power has led to the development of specialized optimization packages like \texttt{opt\_einsum} \cite{opt_einsum}, which focuses on finding the most computationally efficient contraction order for a given expression. The notation has proven particularly valuable in implementing attention mechanisms in transformers \cite{vaswani2023attentionneed}, where complex batched matrix operations can be expressed more clearly than with traditional BLAS calls. While einsum implementations typically compile down to optimized GEMM routines, the notation itself serves as a powerful intermediate representation. Libraries such as TensorFlow \cite{tensorflow} use einsum expressions internally to represent and optimize operations before lowering to platform-specific kernels. This dual nature of einsum as both a user-facing API and an optimization target makes it uniquely suited for bridging high-level tensor expressions with low-level execution strategies. EinHops leverages this same expressive power to provide a clean abstraction for homomorphic tensor operations. 

\section{Conclusion}

In this work, we decomposed the expressive einsum notation into a series of FHE-friendly operations and built EinHops, a system for performing high-dimensional tensor operations in FHE. By mapping the semantics of einsum to a sequence of permutations (via BSGS), broadcasts, and reductions (via rotate-and-sum), EinHops brings a familiar and highly expressive programming model to the constrained environment of RNS-CKKS.

Unlike graph-based compilers that abstract away implementation details behind complex optimization passes, EinHops makes the data layout and operational plan explicit through the einsum string itself. This transparency empowers researchers to reason clearly about the underlying homomorphic operations, how slots are being used, and apply optimizations at each stage of the pipeline. We evaluate EinHops on a variety of tensor operations, from simple transposes to multi-head attention scores used in transformers, demonstrating that einsum notation provides an intuitive yet powerful abstraction for FHE tensor operations.

We embraced minimalism and simplicity throughout this project. We hope our design philosophy makes learning about FHE more accessible and helps practitioners better explore system-level nuances of running FHE programs.

\section*{Acknowledgements}
\noindent
This work was supported in part by Graduate Assistance in Areas of National Need (GAANN). The research was developed with funding from the NSF CAREER award \#2340137 and DARPA, under the Data Protection in Virtual Environments (DPRIVE) program, contract HR0011-21-9-0003. Reagen and Ebel received a gift award from Google. The views, opinions, and/or findings expressed are those of the authors and do not necessarily reflect the views of sponsors.


\bibliographystyle{ACM-Reference-Format}
\bibliography{ccs-sample}


\begin{thebibliography}{00}


\ifx \showCODEN    \undefined \def \showCODEN     #1{\unskip}     \fi
\ifx \showDOI      \undefined \def \showDOI       #1{#1}\fi
\ifx \showISBNx    \undefined \def \showISBNx     #1{\unskip}     \fi
\ifx \showISBNxiii \undefined \def \showISBNxiii  #1{\unskip}     \fi
\ifx \showISSN     \undefined \def \showISSN      #1{\unskip}     \fi
\ifx \showLCCN     \undefined \def \showLCCN      #1{\unskip}     \fi
\ifx \shownote     \undefined \def \shownote      #1{#1}          \fi
\ifx \showarticletitle \undefined \def \showarticletitle #1{#1}   \fi
\ifx \showURL      \undefined \def \showURL       {\relax}        \fi
\providecommand\bibfield[2]{#2}
\providecommand\bibinfo[2]{#2}
\providecommand\natexlab[1]{#1}
\providecommand\showeprint[2][]{arXiv:#2}

\bibitem[\protect\citeauthoryear{??}{pyt}{2025}]%
        {pytorch_stride}
 \bibinfo{year}{2025}\natexlab{}.
\newblock \bibinfo{title}{torch.Tensor.stride — PyTorch Documentation}.
\newblock \bibinfo{howpublished}{\url{https://docs.pytorch.org/docs/stable/generated/torch.Tensor.stride.html}}.   (\bibinfo{year}{2025}).
\newblock
\newblock
\shownote{Accessed: 2025-06-26.}


\bibitem[\protect\citeauthoryear{a.~Smith and Gray}{a.~Smith and Gray}{2018}]%
        {opt_einsum}
\bibfield{author}{\bibinfo{person}{Daniel~G. a. Smith} {and} \bibinfo{person}{Johnnie Gray}.} \bibinfo{year}{2018}\natexlab{}.
\newblock \showarticletitle{opt\_einsum - A Python package for optimizing contraction order for einsum-like expressions}.
\newblock \bibinfo{journal}{{\em Journal of Open Source Software\/}} \bibinfo{volume}{3}, \bibinfo{number}{26} (\bibinfo{year}{2018}), \bibinfo{pages}{753}.
\newblock
\showDOI{%
\url{https://doi.org/10.21105/joss.00753}}


\bibitem[\protect\citeauthoryear{Abadi, Agarwal, Barham, Brevdo, Chen, Citro, Corrado, Davis, Dean, Devin, Ghemawat, Goodfellow, Harp, Irving, Isard, Jia, Jozefowicz, Kaiser, Kudlur, Levenberg, Man\'{e}, Monga, Moore, Murray, Olah, Schuster, Shlens, Steiner, Sutskever, Talwar, Tucker, Vanhoucke, Vasudevan, Vi\'{e}gas, Vinyals, Warden, Wattenberg, Wicke, Yu, and Zheng}{Abadi et~al\mbox{.}}{2015}]%
        {tensorflow}
\bibfield{author}{\bibinfo{person}{Mart\'{i}n Abadi}, \bibinfo{person}{Ashish Agarwal}, \bibinfo{person}{Paul Barham}, \bibinfo{person}{Eugene Brevdo}, \bibinfo{person}{Zhifeng Chen}, \bibinfo{person}{Craig Citro}, \bibinfo{person}{Greg~S. Corrado}, \bibinfo{person}{Andy Davis}, \bibinfo{person}{Jeffrey Dean}, \bibinfo{person}{Matthieu Devin}, \bibinfo{person}{Sanjay Ghemawat}, \bibinfo{person}{Ian Goodfellow}, \bibinfo{person}{Andrew Harp}, \bibinfo{person}{Geoffrey Irving}, \bibinfo{person}{Michael Isard}, \bibinfo{person}{Yangqing Jia}, \bibinfo{person}{Rafal Jozefowicz}, \bibinfo{person}{Lukasz Kaiser}, \bibinfo{person}{Manjunath Kudlur}, \bibinfo{person}{Josh Levenberg}, \bibinfo{person}{Dandelion Man\'{e}}, \bibinfo{person}{Rajat Monga}, \bibinfo{person}{Sherry Moore}, \bibinfo{person}{Derek Murray}, \bibinfo{person}{Chris Olah}, \bibinfo{person}{Mike Schuster}, \bibinfo{person}{Jonathon Shlens}, \bibinfo{person}{Benoit Steiner}, \bibinfo{person}{Ilya Sutskever}, \bibinfo{person}{Kunal Talwar},
  \bibinfo{person}{Paul Tucker}, \bibinfo{person}{Vincent Vanhoucke}, \bibinfo{person}{Vijay Vasudevan}, \bibinfo{person}{Fernanda Vi\'{e}gas}, \bibinfo{person}{Oriol Vinyals}, \bibinfo{person}{Pete Warden}, \bibinfo{person}{Martin Wattenberg}, \bibinfo{person}{Martin Wicke}, \bibinfo{person}{Yuan Yu}, {and} \bibinfo{person}{Xiaoqiang Zheng}.} \bibinfo{year}{2015}\natexlab{}.
\newblock \bibinfo{title}{{TensorFlow}: Large-Scale Machine Learning on Heterogeneous Systems}.
\newblock   (\bibinfo{year}{2015}).
\newblock
\showURL{%
\url{https://www.tensorflow.org/}}
\newblock
\shownote{Software available from tensorflow.org.}


\bibitem[\protect\citeauthoryear{Aharoni, Adir, Baruch, Drucker, Ezov, Farkash, Greenberg, Masalha, Moshkowich, Murik, Shaul, and Soceanu}{Aharoni et~al\mbox{.}}{2023}]%
        {helayers}
\bibfield{author}{\bibinfo{person}{Ehud Aharoni}, \bibinfo{person}{Allon Adir}, \bibinfo{person}{Moran Baruch}, \bibinfo{person}{Nir Drucker}, \bibinfo{person}{Gilad Ezov}, \bibinfo{person}{Ariel Farkash}, \bibinfo{person}{Lev Greenberg}, \bibinfo{person}{Ramy Masalha}, \bibinfo{person}{Guy Moshkowich}, \bibinfo{person}{Dov Murik}, \bibinfo{person}{Hayim Shaul}, {and} \bibinfo{person}{Omri Soceanu}.} \bibinfo{year}{2023}\natexlab{}.
\newblock \showarticletitle{HeLayers: A Tile Tensors Framework for Large Neural Networks on Encrypted Data}.
\newblock \bibinfo{journal}{{\em Proceedings on Privacy Enhancing Technologies\/}} \bibinfo{volume}{2023}, \bibinfo{number}{1} (\bibinfo{date}{Jan.} \bibinfo{year}{2023}), \bibinfo{pages}{325–342}.
\newblock
\showISSN{2299-0984}
\showDOI{%
\url{https://doi.org/10.56553/popets-2023-0020}}


\bibitem[\protect\citeauthoryear{Ansel, Yang, He, Gimelshein, Jain, Voznesensky, Bao, Bell, Berard, Burovski, Chauhan, Chourdia, Constable, Desmaison, DeVito, Ellison, Feng, Gong, Gschwind, Hirsh, Huang, Kalambarkar, Kirsch, Lazos, Lezcano, Liang, Liang, Lu, Luk, Maher, Pan, Puhrsch, Reso, Saroufim, Siraichi, Suk, Suo, Tillet, Wang, Wang, Wen, Zhang, Zhao, Zhou, Zou, Mathews, Chanan, Wu, and Chintala}{Ansel et~al\mbox{.}}{2024}]%
        {pytorch}
\bibfield{author}{\bibinfo{person}{Jason Ansel}, \bibinfo{person}{Edward Yang}, \bibinfo{person}{Horace He}, \bibinfo{person}{Natalia Gimelshein}, \bibinfo{person}{Animesh Jain}, \bibinfo{person}{Michael Voznesensky}, \bibinfo{person}{Bin Bao}, \bibinfo{person}{Peter Bell}, \bibinfo{person}{David Berard}, \bibinfo{person}{Evgeni Burovski}, \bibinfo{person}{Geeta Chauhan}, \bibinfo{person}{Anjali Chourdia}, \bibinfo{person}{Will Constable}, \bibinfo{person}{Alban Desmaison}, \bibinfo{person}{Zachary DeVito}, \bibinfo{person}{Elias Ellison}, \bibinfo{person}{Will Feng}, \bibinfo{person}{Jiong Gong}, \bibinfo{person}{Michael Gschwind}, \bibinfo{person}{Brian Hirsh}, \bibinfo{person}{Sherlock Huang}, \bibinfo{person}{Kshiteej Kalambarkar}, \bibinfo{person}{Laurent Kirsch}, \bibinfo{person}{Michael Lazos}, \bibinfo{person}{Mario Lezcano}, \bibinfo{person}{Yanbo Liang}, \bibinfo{person}{Jason Liang}, \bibinfo{person}{Yinghai Lu}, \bibinfo{person}{CK Luk}, \bibinfo{person}{Bert Maher}, \bibinfo{person}{Yunjie Pan},
  \bibinfo{person}{Christiao Puhrsch}, \bibinfo{person}{Matthias Reso}, \bibinfo{person}{Mark Saroufim}, \bibinfo{person}{Marcos~Yukio Siraichi}, \bibinfo{person}{Helen Suk}, \bibinfo{person}{Michael Suo}, \bibinfo{person}{Phil Tillet}, \bibinfo{person}{Eikan Wang}, \bibinfo{person}{Xiaodong Wang}, \bibinfo{person}{William Wen}, \bibinfo{person}{Shunting Zhang}, \bibinfo{person}{Xu Zhao}, \bibinfo{person}{Keren Zhou}, \bibinfo{person}{Richard Zou}, \bibinfo{person}{Ajit Mathews}, \bibinfo{person}{Gregory Chanan}, \bibinfo{person}{Peng Wu}, {and} \bibinfo{person}{Soumith Chintala}.} \bibinfo{year}{2024}\natexlab{}.
\newblock \showarticletitle{{PyTorch 2: Faster Machine Learning Through Dynamic Python Bytecode Transformation and Graph Compilation}}. In \bibinfo{booktitle}{{\em 29th ACM International Conference on Architectural Support for Programming Languages and Operating Systems, Volume 2 (ASPLOS '24)}}. \bibinfo{publisher}{ACM}.
\newblock
\showDOI{%
\url{https://doi.org/10.1145/3620665.3640366}}


\bibitem[\protect\citeauthoryear{Augonnet, Alexandrescu, Sidelnik, and Garland}{Augonnet et~al\mbox{.}}{2024}]%
        {cudastf}
\bibfield{author}{\bibinfo{person}{C\'{e}dric Augonnet}, \bibinfo{person}{Andrei Alexandrescu}, \bibinfo{person}{Albert Sidelnik}, {and} \bibinfo{person}{Michael Garland}.} \bibinfo{year}{2024}\natexlab{}.
\newblock \showarticletitle{CUDASTF: Bridging the Gap Between CUDA and Task Parallelism}. In \bibinfo{booktitle}{{\em Proceedings of the International Conference for High Performance Computing, Networking, Storage, and Analysis}} {\em (\bibinfo{series}{SC '24})}. \bibinfo{publisher}{IEEE Press}, Article \bibinfo{articleno}{43}, \bibinfo{numpages}{17}~pages.
\newblock
\showISBNx{9798350352917}
\showDOI{%
\url{https://doi.org/10.1109/SC41406.2024.00049}}


\bibitem[\protect\citeauthoryear{Blacher, Staudt, Klaus, Wenig, Merk, Breuer, Engel, Laue, and Giesen}{Blacher et~al\mbox{.}}{2024}]%
        {einsum_bench}
\bibfield{author}{\bibinfo{person}{Mark Blacher}, \bibinfo{person}{Christoph Staudt}, \bibinfo{person}{Julien Klaus}, \bibinfo{person}{Maurice Wenig}, \bibinfo{person}{Niklas Merk}, \bibinfo{person}{Alexander Breuer}, \bibinfo{person}{Max Engel}, \bibinfo{person}{S\"{o}ren Laue}, {and} \bibinfo{person}{Joachim Giesen}.} \bibinfo{year}{2024}\natexlab{}.
\newblock \showarticletitle{Einsum Benchmark: Enabling the Development of Next-Generation Tensor Execution Engines}. In \bibinfo{booktitle}{{\em Advances in Neural Information Processing Systems}}, \bibfield{editor}{\bibinfo{person}{A.~Globerson}, \bibinfo{person}{L.~Mackey}, \bibinfo{person}{D.~Belgrave}, \bibinfo{person}{A.~Fan}, \bibinfo{person}{U.~Paquet}, \bibinfo{person}{J.~Tomczak}, {and} \bibinfo{person}{C.~Zhang}} (Eds.), Vol.~\bibinfo{volume}{37}. \bibinfo{publisher}{Curran Associates, Inc.}, \bibinfo{pages}{98033--98048}.
\newblock
\showURL{%
\url{https://proceedings.neurips.cc/paper_files/paper/2024/file/b1bbfdb9197bfc819a52c34dce493f85-Paper-Datasets_and_Benchmarks_Track.pdf}}


\bibitem[\protect\citeauthoryear{Bradbury, Frostig, Hawkins, Johnson, Leary, Maclaurin, Necula, Paszke, Vander{P}las, Wanderman-{M}ilne, and Zhang}{Bradbury et~al\mbox{.}}{2018}]%
        {jax}
\bibfield{author}{\bibinfo{person}{James Bradbury}, \bibinfo{person}{Roy Frostig}, \bibinfo{person}{Peter Hawkins}, \bibinfo{person}{Matthew~James Johnson}, \bibinfo{person}{Chris Leary}, \bibinfo{person}{Dougal Maclaurin}, \bibinfo{person}{George Necula}, \bibinfo{person}{Adam Paszke}, \bibinfo{person}{Jake Vander{P}las}, \bibinfo{person}{Skye Wanderman-{M}ilne}, {and} \bibinfo{person}{Qiao Zhang}.} \bibinfo{year}{2018}\natexlab{}.
\newblock \bibinfo{title}{{JAX}: composable transformations of {P}ython+{N}um{P}y programs}.
\newblock   (\bibinfo{year}{2018}).
\newblock
\showURL{%
\url{http://github.com/jax-ml/jax}}


\bibitem[\protect\citeauthoryear{Brakerski}{Brakerski}{2012}]%
        {b}
\bibfield{author}{\bibinfo{person}{Zvika Brakerski}.} \bibinfo{year}{2012}\natexlab{}.
\newblock \bibinfo{title}{Fully Homomorphic Encryption without Modulus Switching from Classical {GapSVP}}.
\newblock \bibinfo{howpublished}{Cryptology {ePrint} Archive, Paper 2012/078}.   (\bibinfo{year}{2012}).
\newblock
\showURL{%
\url{https://eprint.iacr.org/2012/078}}


\bibitem[\protect\citeauthoryear{Brakerski, Gentry, and Vaikuntanathan}{Brakerski et~al\mbox{.}}{2011}]%
        {bgv}
\bibfield{author}{\bibinfo{person}{Zvika Brakerski}, \bibinfo{person}{Craig Gentry}, {and} \bibinfo{person}{Vinod Vaikuntanathan}.} \bibinfo{year}{2011}\natexlab{}.
\newblock \bibinfo{title}{Fully Homomorphic Encryption without Bootstrapping}.
\newblock \bibinfo{howpublished}{Cryptology {ePrint} Archive, Paper 2011/277}.   (\bibinfo{year}{2011}).
\newblock
\showURL{%
\url{https://eprint.iacr.org/2011/277}}


\bibitem[\protect\citeauthoryear{Cheon, Han, Kim, Kim, and Song}{Cheon et~al\mbox{.}}{2018a}]%
        {bootstrapping}
\bibfield{author}{\bibinfo{person}{Jung~Hee Cheon}, \bibinfo{person}{Kyoohyung Han}, \bibinfo{person}{Andrey Kim}, \bibinfo{person}{Miran Kim}, {and} \bibinfo{person}{Yongsoo Song}.} \bibinfo{year}{2018}\natexlab{a}.
\newblock \bibinfo{title}{Bootstrapping for Approximate Homomorphic Encryption}.
\newblock \bibinfo{howpublished}{Cryptology {ePrint} Archive, Paper 2018/153}.   (\bibinfo{year}{2018}).
\newblock
\showURL{%
\url{https://eprint.iacr.org/2018/153}}


\bibitem[\protect\citeauthoryear{Cheon, Han, Kim, Kim, and Song}{Cheon et~al\mbox{.}}{2018b}]%
        {rns-ckks}
\bibfield{author}{\bibinfo{person}{Jung~Hee Cheon}, \bibinfo{person}{Kyoohyung Han}, \bibinfo{person}{Andrey Kim}, \bibinfo{person}{Miran Kim}, {and} \bibinfo{person}{Yongsoo Song}.} \bibinfo{year}{2018}\natexlab{b}.
\newblock \bibinfo{title}{A Full {RNS} Variant of Approximate Homomorphic Encryption}.
\newblock \bibinfo{howpublished}{Cryptology {ePrint} Archive, Paper 2018/931}.   (\bibinfo{year}{2018}).
\newblock
\showURL{%
\url{https://eprint.iacr.org/2018/931}}


\bibitem[\protect\citeauthoryear{Cheon, Kim, Kim, and Song}{Cheon et~al\mbox{.}}{2016}]%
        {ckks}
\bibfield{author}{\bibinfo{person}{Jung~Hee Cheon}, \bibinfo{person}{Andrey Kim}, \bibinfo{person}{Miran Kim}, {and} \bibinfo{person}{Yongsoo Song}.} \bibinfo{year}{2016}\natexlab{}.
\newblock \bibinfo{title}{Homomorphic Encryption for Arithmetic of Approximate Numbers}.
\newblock \bibinfo{howpublished}{Cryptology {ePrint} Archive, Paper 2016/421}.   (\bibinfo{year}{2016}).
\newblock
\showURL{%
\url{https://eprint.iacr.org/2016/421}}


\bibitem[\protect\citeauthoryear{Cheon, Lee, Kim, Lee, Jung, Kim, Lee, and Kim}{Cheon et~al\mbox{.}}{2024}]%
        {dacapo}
\bibfield{author}{\bibinfo{person}{Seonyoung Cheon}, \bibinfo{person}{Yongwoo Lee}, \bibinfo{person}{Dongkwan Kim}, \bibinfo{person}{Ju~Min Lee}, \bibinfo{person}{Sunchul Jung}, \bibinfo{person}{Taekyung Kim}, \bibinfo{person}{Dongyoon Lee}, {and} \bibinfo{person}{Hanjun Kim}.} \bibinfo{year}{2024}\natexlab{}.
\newblock \showarticletitle{{DaCapo}: Automatic Bootstrapping Management for Efficient Fully Homomorphic Encryption}. In \bibinfo{booktitle}{{\em 33rd USENIX Security Symposium (USENIX Security 24)}}. \bibinfo{publisher}{USENIX Association}, \bibinfo{address}{Philadelphia, PA}, \bibinfo{pages}{6993--7010}.
\newblock
\showISBNx{978-1-939133-44-1}
\showURL{%
\url{https://www.usenix.org/conference/usenixsecurity24/presentation/cheon}}


\bibitem[\protect\citeauthoryear{Chetlur, Woolley, Vandermersch, Cohen, Tran, Catanzaro, and Shelhamer}{Chetlur et~al\mbox{.}}{2014}]%
        {cudnn}
\bibfield{author}{\bibinfo{person}{Sharan Chetlur}, \bibinfo{person}{Cliff Woolley}, \bibinfo{person}{Philippe Vandermersch}, \bibinfo{person}{Jonathan Cohen}, \bibinfo{person}{John Tran}, \bibinfo{person}{Bryan Catanzaro}, {and} \bibinfo{person}{Evan Shelhamer}.} \bibinfo{year}{2014}\natexlab{}.
\newblock \bibinfo{title}{cuDNN: Efficient Primitives for Deep Learning}.
\newblock   (\bibinfo{year}{2014}).
\newblock
\showeprint[arxiv]{cs.NE/1410.0759}
\showURL{%
\url{https://arxiv.org/abs/1410.0759}}


\bibitem[\protect\citeauthoryear{Chillotti, Gama, Georgieva, and Izabachène}{Chillotti et~al\mbox{.}}{2018}]%
        {tfhe}
\bibfield{author}{\bibinfo{person}{Ilaria Chillotti}, \bibinfo{person}{Nicolas Gama}, \bibinfo{person}{Mariya Georgieva}, {and} \bibinfo{person}{Malika Izabachène}.} \bibinfo{year}{2018}\natexlab{}.
\newblock \bibinfo{title}{{TFHE}: Fast Fully Homomorphic Encryption over the Torus}.
\newblock \bibinfo{howpublished}{Cryptology {ePrint} Archive, Paper 2018/421}.   (\bibinfo{year}{2018}).
\newblock
\showURL{%
\url{https://eprint.iacr.org/2018/421}}


\bibitem[\protect\citeauthoryear{Contributors}{Contributors}{2024}]%
        {pytorch_broadcasting}
\bibfield{author}{\bibinfo{person}{PyTorch Contributors}.} \bibinfo{year}{2024}\natexlab{}.
\newblock \bibinfo{title}{Broadcasting semantics}.
\newblock \bibinfo{howpublished}{\url{https://docs.pytorch.org/docs/stable/notes/broadcasting.html}}.   (\bibinfo{year}{2024}).
\newblock
\newblock
\shownote{Accessed: 2025-06-26.}


\bibitem[\protect\citeauthoryear{Cowan, Dangwal, Alaghi, Trippel, Lee, and Reagen}{Cowan et~al\mbox{.}}{2021}]%
        {porcupine}
\bibfield{author}{\bibinfo{person}{Meghan Cowan}, \bibinfo{person}{Deeksha Dangwal}, \bibinfo{person}{Armin Alaghi}, \bibinfo{person}{Caroline Trippel}, \bibinfo{person}{Vincent~T. Lee}, {and} \bibinfo{person}{Brandon Reagen}.} \bibinfo{year}{2021}\natexlab{}.
\newblock \showarticletitle{Porcupine: a synthesizing compiler for vectorized homomorphic encryption}. In \bibinfo{booktitle}{{\em Proceedings of the 42nd ACM SIGPLAN International Conference on Programming Language Design and Implementation}} {\em (\bibinfo{series}{PLDI 2021})}. \bibinfo{publisher}{Association for Computing Machinery}, \bibinfo{address}{New York, NY, USA}, \bibinfo{pages}{375–389}.
\newblock
\showISBNx{9781450383912}
\showDOI{%
\url{https://doi.org/10.1145/3453483.3454050}}


\bibitem[\protect\citeauthoryear{Dathathri, Kostova, Saarikivi, Dai, Laine, and Musuvathi}{Dathathri et~al\mbox{.}}{2020}]%
        {eva}
\bibfield{author}{\bibinfo{person}{Roshan Dathathri}, \bibinfo{person}{Blagovesta Kostova}, \bibinfo{person}{Olli Saarikivi}, \bibinfo{person}{Wei Dai}, \bibinfo{person}{Kim Laine}, {and} \bibinfo{person}{Madan Musuvathi}.} \bibinfo{year}{2020}\natexlab{}.
\newblock \showarticletitle{{EVA}: an encrypted vector arithmetic language and compiler for efficient homomorphic computation}. In \bibinfo{booktitle}{{\em Proceedings of the 41st {ACM} {SIGPLAN} Conference on Programming Language Design and Implementation}}. \bibinfo{publisher}{{ACM}}.
\newblock
\showDOI{%
\url{https://doi.org/10.1145/3385412.3386023}}


\bibitem[\protect\citeauthoryear{Dathathri, Saarikivi, Chen, Laine, Lauter, Maleki, Musuvathi, and Mytkowicz}{Dathathri et~al\mbox{.}}{2019}]%
        {chet}
\bibfield{author}{\bibinfo{person}{Roshan Dathathri}, \bibinfo{person}{Olli Saarikivi}, \bibinfo{person}{Hao Chen}, \bibinfo{person}{Kim Laine}, \bibinfo{person}{Kristin Lauter}, \bibinfo{person}{Saeed Maleki}, \bibinfo{person}{Madanlal Musuvathi}, {and} \bibinfo{person}{Todd Mytkowicz}.} \bibinfo{year}{2019}\natexlab{}.
\newblock \showarticletitle{CHET: an optimizing compiler for fully-homomorphic neural-network inferencing}. In \bibinfo{booktitle}{{\em Proceedings of the 40th ACM SIGPLAN Conference on Programming Language Design and Implementation}} {\em (\bibinfo{series}{PLDI 2019})}. \bibinfo{publisher}{Association for Computing Machinery}, \bibinfo{address}{New York, NY, USA}, \bibinfo{pages}{142–156}.
\newblock
\showISBNx{9781450367127}
\showDOI{%
\url{https://doi.org/10.1145/3314221.3314628}}


\bibitem[\protect\citeauthoryear{Del{\'{e}}tang, Ruoss, Duquenne, Catt, Genewein, Mattern, Grau{-}Moya, Wenliang, Aitchison, Orseau, Hutter, and Veness}{Del{\'{e}}tang et~al\mbox{.}}{2024}]%
        {deletang2024language}
\bibfield{author}{\bibinfo{person}{Gr{\'{e}}goire Del{\'{e}}tang}, \bibinfo{person}{Anian Ruoss}, \bibinfo{person}{Paul{-}Ambroise Duquenne}, \bibinfo{person}{Elliot Catt}, \bibinfo{person}{Tim Genewein}, \bibinfo{person}{Christopher Mattern}, \bibinfo{person}{Jordi Grau{-}Moya}, \bibinfo{person}{Li~Kevin Wenliang}, \bibinfo{person}{Matthew Aitchison}, \bibinfo{person}{Laurent Orseau}, \bibinfo{person}{Marcus Hutter}, {and} \bibinfo{person}{Joel Veness}.} \bibinfo{year}{2024}\natexlab{}.
\newblock \showarticletitle{Language Modeling Is Compression}. In \bibinfo{booktitle}{{\em {ICLR}}}.
\newblock


\bibitem[\protect\citeauthoryear{DESILO}{DESILO}{2023}]%
        {Liberate_FHE}
\bibfield{author}{\bibinfo{person}{DESILO}.} \bibinfo{year}{2023}\natexlab{}.
\newblock \bibinfo{title}{{Liberate.FHE: A New FHE Library for Bridging the Gap between Theory and Practice with a Focus on Performance and Accuracy}}.
\newblock   (\bibinfo{year}{2023}).
\newblock
\newblock
\shownote{\url{https://github.com/Desilo/liberate-fhe}.}


\bibitem[\protect\citeauthoryear{Ebel, Garimella, and Reagen}{Ebel et~al\mbox{.}}{2025}]%
        {orion}
\bibfield{author}{\bibinfo{person}{Austin Ebel}, \bibinfo{person}{Karthik Garimella}, {and} \bibinfo{person}{Brandon Reagen}.} \bibinfo{year}{2025}\natexlab{}.
\newblock \showarticletitle{Orion: A Fully Homomorphic Encryption Framework for Deep Learning}. In \bibinfo{booktitle}{{\em Proceedings of the 30th ACM International Conference on Architectural Support for Programming Languages and Operating Systems, Volume 2}} {\em (\bibinfo{series}{ASPLOS '25})}. \bibinfo{publisher}{Association for Computing Machinery}, \bibinfo{address}{New York, NY, USA}, \bibinfo{pages}{734–749}.
\newblock
\showISBNx{9798400710797}
\showDOI{%
\url{https://doi.org/10.1145/3676641.3716008}}


\bibitem[\protect\citeauthoryear{Einstein}{Einstein}{1916}]%
        {einstein}
\bibfield{author}{\bibinfo{person}{Albert Einstein}.} \bibinfo{year}{1916}\natexlab{}.
\newblock \showarticletitle{{The foundation of the general theory of relativity.}}
\newblock \bibinfo{journal}{{\em Annalen Phys.\/}} \bibinfo{volume}{49}, \bibinfo{number}{7} (\bibinfo{year}{1916}), \bibinfo{pages}{769--822}.
\newblock
\showDOI{%
\url{https://doi.org/10.1002/andp.19163540702}}


\bibitem[\protect\citeauthoryear{Fan and Vercauteren}{Fan and Vercauteren}{2012}]%
        {fv}
\bibfield{author}{\bibinfo{person}{Junfeng Fan} {and} \bibinfo{person}{Frederik Vercauteren}.} \bibinfo{year}{2012}\natexlab{}.
\newblock \bibinfo{title}{Somewhat Practical Fully Homomorphic Encryption}.
\newblock \bibinfo{howpublished}{Cryptology {ePrint} Archive, Paper 2012/144}.   (\bibinfo{year}{2012}).
\newblock
\showURL{%
\url{https://eprint.iacr.org/2012/144}}


\bibitem[\protect\citeauthoryear{Garner}{Garner}{1959}]%
        {rns}
\bibfield{author}{\bibinfo{person}{Harvey~L. Garner}.} \bibinfo{year}{1959}\natexlab{}.
\newblock \showarticletitle{The residue number system}. In \bibinfo{booktitle}{{\em Papers Presented at the the March 3-5, 1959, Western Joint Computer Conference}} {\em (\bibinfo{series}{IRE-AIEE-ACM '59 (Western)})}. \bibinfo{publisher}{Association for Computing Machinery}, \bibinfo{address}{New York, NY, USA}, \bibinfo{pages}{146–153}.
\newblock
\showISBNx{9781450378659}
\showDOI{%
\url{https://doi.org/10.1145/1457838.1457864}}


\bibitem[\protect\citeauthoryear{Gentry}{Gentry}{2009}]%
        {fhe}
\bibfield{author}{\bibinfo{person}{Craig Gentry}.} \bibinfo{year}{2009}\natexlab{}.
\newblock \showarticletitle{Fully homomorphic encryption using ideal lattices}. In \bibinfo{booktitle}{{\em Proceedings of the Forty-First Annual ACM Symposium on Theory of Computing}} {\em (\bibinfo{series}{STOC '09})}. \bibinfo{publisher}{Association for Computing Machinery}, \bibinfo{address}{New York, NY, USA}, \bibinfo{pages}{169–178}.
\newblock
\showISBNx{9781605585062}
\showDOI{%
\url{https://doi.org/10.1145/1536414.1536440}}


\bibitem[\protect\citeauthoryear{Halevi and Shoup}{Halevi and Shoup}{2014}]%
        {halevishoup}
\bibfield{author}{\bibinfo{person}{Shai Halevi} {and} \bibinfo{person}{Victor Shoup}.} \bibinfo{year}{2014}\natexlab{}.
\newblock \bibinfo{title}{Algorithms in {HElib}}.
\newblock \bibinfo{howpublished}{Cryptology {ePrint} Archive, Paper 2014/106}.   (\bibinfo{year}{2014}).
\newblock
\showURL{%
\url{https://eprint.iacr.org/2014/106}}


\bibitem[\protect\citeauthoryear{Halevi and Shoup}{Halevi and Shoup}{2018}]%
        {faster_helib}
\bibfield{author}{\bibinfo{person}{Shai Halevi} {and} \bibinfo{person}{Victor Shoup}.} \bibinfo{year}{2018}\natexlab{}.
\newblock \showarticletitle{Faster Homomorphic Linear Transformations in HElib}. In \bibinfo{booktitle}{{\em Advances in Cryptology – CRYPTO 2018: 38th Annual International Cryptology Conference, Santa Barbara, CA, USA, August 19–23, 2018, Proceedings, Part I}}. \bibinfo{publisher}{Springer-Verlag}, \bibinfo{address}{Berlin, Heidelberg}, \bibinfo{pages}{93–120}.
\newblock
\showISBNx{978-3-319-96883-4}
\showDOI{%
\url{https://doi.org/10.1007/978-3-319-96884-1_4}}


\bibitem[\protect\citeauthoryear{Han and Ki}{Han and Ki}{2019}]%
        {keyswitch}
\bibfield{author}{\bibinfo{person}{Kyoohyung Han} {and} \bibinfo{person}{Dohyeong Ki}.} \bibinfo{year}{2019}\natexlab{}.
\newblock \bibinfo{title}{Better Bootstrapping for Approximate Homomorphic Encryption}.
\newblock \bibinfo{howpublished}{Cryptology {ePrint} Archive, Paper 2019/688}.   (\bibinfo{year}{2019}).
\newblock
\showURL{%
\url{https://eprint.iacr.org/2019/688}}


\bibitem[\protect\citeauthoryear{Harris, Millman, van~der Walt, Gommers, Virtanen, Cournapeau, Wieser, Taylor, Berg, Smith, Kern, Picus, Hoyer, van Kerkwijk, Brett, Haldane, del R{\'{i}}o, Wiebe, Peterson, G{\'{e}}rard-Marchant, Sheppard, Reddy, Weckesser, Abbasi, Gohlke, and Oliphant}{Harris et~al\mbox{.}}{2020}]%
        {numpy}
\bibfield{author}{\bibinfo{person}{Charles~R. Harris}, \bibinfo{person}{K.~Jarrod Millman}, \bibinfo{person}{St{\'{e}}fan~J. van~der Walt}, \bibinfo{person}{Ralf Gommers}, \bibinfo{person}{Pauli Virtanen}, \bibinfo{person}{David Cournapeau}, \bibinfo{person}{Eric Wieser}, \bibinfo{person}{Julian Taylor}, \bibinfo{person}{Sebastian Berg}, \bibinfo{person}{Nathaniel~J. Smith}, \bibinfo{person}{Robert Kern}, \bibinfo{person}{Matti Picus}, \bibinfo{person}{Stephan Hoyer}, \bibinfo{person}{Marten~H. van Kerkwijk}, \bibinfo{person}{Matthew Brett}, \bibinfo{person}{Allan Haldane}, \bibinfo{person}{Jaime~Fern{\'{a}}ndez del R{\'{i}}o}, \bibinfo{person}{Mark Wiebe}, \bibinfo{person}{Pearu Peterson}, \bibinfo{person}{Pierre G{\'{e}}rard-Marchant}, \bibinfo{person}{Kevin Sheppard}, \bibinfo{person}{Tyler Reddy}, \bibinfo{person}{Warren Weckesser}, \bibinfo{person}{Hameer Abbasi}, \bibinfo{person}{Christoph Gohlke}, {and} \bibinfo{person}{Travis~E. Oliphant}.} \bibinfo{year}{2020}\natexlab{}.
\newblock \showarticletitle{Array programming with {NumPy}}.
\newblock \bibinfo{journal}{{\em Nature\/}} \bibinfo{volume}{585}, \bibinfo{number}{7825} (\bibinfo{date}{Sept.} \bibinfo{year}{2020}), \bibinfo{pages}{357--362}.
\newblock
\showDOI{%
\url{https://doi.org/10.1038/s41586-020-2649-2}}


\bibitem[\protect\citeauthoryear{Kim and Guyot}{Kim and Guyot}{2023}]%
        {coeff-encoding}
\bibfield{author}{\bibinfo{person}{Dongwoo Kim} {and} \bibinfo{person}{Cyril Guyot}.} \bibinfo{year}{2023}\natexlab{}.
\newblock \showarticletitle{Optimized Privacy-Preserving CNN Inference With Fully Homomorphic Encryption}.
\newblock \bibinfo{journal}{{\em IEEE Transactions on Information Forensics and Security\/}}  \bibinfo{volume}{18} (\bibinfo{year}{2023}), \bibinfo{pages}{2175--2187}.
\newblock
\showDOI{%
\url{https://doi.org/10.1109/TIFS.2023.3263631}}


\bibitem[\protect\citeauthoryear{Krastev, Samardzic, Langowski, Devadas, and Sanchez}{Krastev et~al\mbox{.}}{2024}]%
        {fhelipe}
\bibfield{author}{\bibinfo{person}{Aleksandar Krastev}, \bibinfo{person}{Nikola Samardzic}, \bibinfo{person}{Simon Langowski}, \bibinfo{person}{Srinivas Devadas}, {and} \bibinfo{person}{Daniel Sanchez}.} \bibinfo{year}{2024}\natexlab{}.
\newblock \showarticletitle{A Tensor Compiler with Automatic Data Packing for Simple and Efficient Fully Homomorphic Encryption}.
\newblock \bibinfo{journal}{{\em Proc. ACM Program. Lang.\/}} \bibinfo{volume}{8}, \bibinfo{number}{PLDI}, Article \bibinfo{articleno}{152} (\bibinfo{date}{June} \bibinfo{year}{2024}), \bibinfo{numpages}{25}~pages.
\newblock
\showDOI{%
\url{https://doi.org/10.1145/3656382}}


\bibitem[\protect\citeauthoryear{Kun}{Kun}{2024}]%
        {kun2024shiftnetworks}
\bibfield{author}{\bibinfo{person}{Jeremy Kun}.} \bibinfo{year}{2024}\natexlab{}.
\newblock \bibinfo{title}{Shift Networks}.
\newblock   (\bibinfo{date}{Sept.~2} \bibinfo{year}{2024}).
\newblock
\showURL{%
\url{https://www.jeremykun.com/2024/09/02/shift-networks/}}
\newblock
\shownote{Math and Programming blog post.}


\bibitem[\protect\citeauthoryear{Lawson, Hanson, Kincaid, and Krogh}{Lawson et~al\mbox{.}}{1979}]%
        {BLAS}
\bibfield{author}{\bibinfo{person}{C. Lawson}, \bibinfo{person}{Richard Hanson}, \bibinfo{person}{David Kincaid}, {and} \bibinfo{person}{Fred Krogh}.} \bibinfo{year}{1979}\natexlab{}.
\newblock \showarticletitle{Basic linear algebra subprograms for FORTRAN usage}.
\newblock \bibinfo{journal}{{\em ACM Trans. Math. Softw.\/}}  \bibinfo{volume}{5} (\bibinfo{date}{09} \bibinfo{year}{1979}), \bibinfo{pages}{308--323}.
\newblock
\showDOI{%
\url{https://doi.org/10.1145/355841.355847}}


\bibitem[\protect\citeauthoryear{Lee, Lee, Lee, Kim, Kim, No, and Choi}{Lee et~al\mbox{.}}{2022b}]%
        {lee2022}
\bibfield{author}{\bibinfo{person}{Eunsang Lee}, \bibinfo{person}{Joon-Woo Lee}, \bibinfo{person}{Junghyun Lee}, \bibinfo{person}{Young-Sik Kim}, \bibinfo{person}{Yongjune Kim}, \bibinfo{person}{Jong-Seon No}, {and} \bibinfo{person}{Woosuk Choi}.} \bibinfo{year}{2022}\natexlab{b}.
\newblock \showarticletitle{Low-Complexity Deep Convolutional Neural Networks on Fully Homomorphic Encryption Using Multiplexed Parallel Convolutions}. In \bibinfo{booktitle}{{\em Proceedings of the 39th International Conference on Machine Learning}} {\em (\bibinfo{series}{Proceedings of Machine Learning Research})}, \bibfield{editor}{\bibinfo{person}{Kamalika Chaudhuri}, \bibinfo{person}{Stefanie Jegelka}, \bibinfo{person}{Le~Song}, \bibinfo{person}{Csaba Szepesvari}, \bibinfo{person}{Gang Niu}, {and} \bibinfo{person}{Sivan Sabato}} (Eds.), Vol.~\bibinfo{volume}{162}. \bibinfo{publisher}{PMLR}, \bibinfo{pages}{12403--12422}.
\newblock
\showURL{%
\url{https://proceedings.mlr.press/v162/lee22e.html}}


\bibitem[\protect\citeauthoryear{Lee, Heo, Cheon, Jeong, Kim, Kim, Lee, and Kim}{Lee et~al\mbox{.}}{2022a}]%
        {hecate}
\bibfield{author}{\bibinfo{person}{Yongwoo Lee}, \bibinfo{person}{Seonyeong Heo}, \bibinfo{person}{Seonyoung Cheon}, \bibinfo{person}{Shinnung Jeong}, \bibinfo{person}{Changsu Kim}, \bibinfo{person}{Eunkyung Kim}, \bibinfo{person}{Dongyoon Lee}, {and} \bibinfo{person}{Hanjun Kim}.} \bibinfo{year}{2022}\natexlab{a}.
\newblock \showarticletitle{HECATE: Performance-Aware Scale Optimization for Homomorphic Encryption Compiler}. In \bibinfo{booktitle}{{\em 2022 IEEE/ACM International Symposium on Code Generation and Optimization (CGO)}}. \bibinfo{pages}{193--204}.
\newblock
\showDOI{%
\url{https://doi.org/10.1109/CGO53902.2022.9741265}}


\bibitem[\protect\citeauthoryear{Moon, Yoo, Jiang, and Kim}{Moon et~al\mbox{.}}{2024}]%
        {thor}
\bibfield{author}{\bibinfo{person}{Jungho Moon}, \bibinfo{person}{Dongwoo Yoo}, \bibinfo{person}{Xiaoqian Jiang}, {and} \bibinfo{person}{Miran Kim}.} \bibinfo{year}{2024}\natexlab{}.
\newblock \bibinfo{title}{{THOR}: Secure Transformer Inference with Homomorphic Encryption}.
\newblock \bibinfo{howpublished}{Cryptology {ePrint} Archive, Paper 2024/1881}.   (\bibinfo{year}{2024}).
\newblock
\showURL{%
\url{https://eprint.iacr.org/2024/1881}}


\bibitem[\protect\citeauthoryear{{NVIDIA Corporation}}{{NVIDIA Corporation}}{2025}]%
        {cuBLAS}
\bibfield{author}{\bibinfo{person}{{NVIDIA Corporation}}.} \bibinfo{year}{2025}\natexlab{}.
\newblock \bibinfo{booktitle}{{\em cuBLAS Library}}.
\newblock NVIDIA Corporation.
\newblock
\newblock
\shownote{Version 12.9. Documentation: \url{https://docs.nvidia.com/cuda/cublas/}.}


\bibitem[\protect\citeauthoryear{{PyTorch Team}}{{PyTorch Team}}{2025}]%
        {pytorch-einsum}
\bibfield{author}{\bibinfo{person}{{PyTorch Team}}.} \bibinfo{year}{2025}\natexlab{}.
\newblock \bibinfo{title}{torch.einsum}.
\newblock   (\bibinfo{year}{2025}).
\newblock
\showURL{%
\url{https://pytorch.org/docs/stable/generated/torch.einsum.html}}
\newblock
\shownote{Accessed: 2025-06-21.}


\bibitem[\protect\citeauthoryear{Rockt{\"a}schel}{Rockt{\"a}schel}{2018}]%
        {rocktaschel2018einsum}
\bibfield{author}{\bibinfo{person}{Tim Rockt{\"a}schel}.} \bibinfo{year}{2018}\natexlab{}.
\newblock \bibinfo{title}{Einsum is All You Need -- Einstein Summation in Deep Learning}.
\newblock   (\bibinfo{year}{2018}).
\newblock
\showURL{%
\url{https://web.archive.org/web/20250514130020/https://rockt.ai/2018/04/30/einsum}}
\newblock
\shownote{Accessed: 2025-06-25.}


\bibitem[\protect\citeauthoryear{Vaswani, Shazeer, Parmar, Uszkoreit, Jones, Gomez, Kaiser, and Polosukhin}{Vaswani et~al\mbox{.}}{2023}]%
        {vaswani2023attentionneed}
\bibfield{author}{\bibinfo{person}{Ashish Vaswani}, \bibinfo{person}{Noam Shazeer}, \bibinfo{person}{Niki Parmar}, \bibinfo{person}{Jakob Uszkoreit}, \bibinfo{person}{Llion Jones}, \bibinfo{person}{Aidan~N. Gomez}, \bibinfo{person}{Lukasz Kaiser}, {and} \bibinfo{person}{Illia Polosukhin}.} \bibinfo{year}{2023}\natexlab{}.
\newblock \bibinfo{title}{Attention Is All You Need}.
\newblock   (\bibinfo{year}{2023}).
\newblock
\showeprint[arxiv]{cs.CL/1706.03762}
\showURL{%
\url{https://arxiv.org/abs/1706.03762}}


\bibitem[\protect\citeauthoryear{Viand, Jattke, Haller, and Hithnawi}{Viand et~al\mbox{.}}{2023}]%
        {heco}
\bibfield{author}{\bibinfo{person}{Alexander Viand}, \bibinfo{person}{Patrick Jattke}, \bibinfo{person}{Miro Haller}, {and} \bibinfo{person}{Anwar Hithnawi}.} \bibinfo{year}{2023}\natexlab{}.
\newblock \showarticletitle{HECO: fully homomorphic encryption compiler}. In \bibinfo{booktitle}{{\em Proceedings of the 32nd USENIX Conference on Security Symposium}} {\em (\bibinfo{series}{SEC '23})}. \bibinfo{publisher}{USENIX Association}, \bibinfo{address}{USA}, Article \bibinfo{articleno}{264}, \bibinfo{numpages}{18}~pages.
\newblock
\showISBNx{978-1-939133-37-3}


\bibitem[\protect\citeauthoryear{Vos, Vos, and Erkin}{Vos et~al\mbox{.}}{2022}]%
        {vve}
\bibfield{author}{\bibinfo{person}{Jelle Vos}, \bibinfo{person}{Dani{\"e}l Vos}, {and} \bibinfo{person}{Zekeriya Erkin}.} \bibinfo{year}{2022}\natexlab{}.
\newblock \showarticletitle{Efficient Circuits for Permuting and Mapping Packed Values Across Leveled Homomorphic Ciphertexts}. In \bibinfo{booktitle}{{\em Computer Security -- ESORICS 2022}}, \bibfield{editor}{\bibinfo{person}{Vijayalakshmi Atluri}, \bibinfo{person}{Roberto Di~Pietro}, \bibinfo{person}{Christian~D. Jensen}, {and} \bibinfo{person}{Weizhi Meng}} (Eds.). \bibinfo{publisher}{Springer International Publishing}, \bibinfo{address}{Cham}, \bibinfo{pages}{408--423}.
\newblock
\showISBNx{978-3-031-17140-6}


\bibitem[\protect\citeauthoryear{Yang}{Yang}{2019}]%
        {yang2019pytorchinternals}
\bibfield{author}{\bibinfo{person}{Edward~Z. Yang}.} \bibinfo{year}{2019}\natexlab{}.
\newblock \bibinfo{title}{{PyTorch Internals}}.
\newblock \bibinfo{howpublished}{\url{https://blog.ezyang.com/2019/05/pytorch-internals/}}.   (\bibinfo{date}{May} \bibinfo{year}{2019}).
\newblock
\showURL{%
\url{https://blog.ezyang.com/2019/05/pytorch-internals/}}
\newblock
\shownote{Long‑form essay based on a talk at the PyTorch NYC meetup (May 14, 2019).}


\end{thebibliography}

\end{document}